\documentclass{aa}
\usepackage{graphicx}
\usepackage{txfonts}
\usepackage[pdftitle={Optimized transit-fitting algorithm to search for periodic transits of small planets}, colorlinks = true, breaklinks = true, citecolor = blue, linkcolor = blue, urlcolor = blue, pdfauthor = {Hippke and Heller}]{hyperref}

\begin{document}
\title{Optimized transit detection algorithm to search for\\periodic transits of small planets}
\author{Michael Hippke\inst{1} \and Ren\'{e} Heller\inst{2}}

\institute{Sonneberg Observatory, Sternwartestr. 32, 96515 Sonneberg, Germany; \href{mailto:michael@hippke.org}{michael@hippke.org}
\and Max Planck Institute for Solar System Research, Justus-von-Liebig-Weg 3, 37077 G\"ottingen, Germany; \href{mailto:heller@mps.mpg.de}{heller@mps.mpg.de}}

\date{Accepted 22 January 2019}

\abstract{We present a new method to detect planetary transits from time-series photometry, the transit least squares ({\tt TLS}) algorithm. {\tt TLS} searches for transit-like features while taking the stellar limb darkening and planetary ingress and egress into account. We have optimized {\tt TLS} for both signal detection efficiency (SDE) of small planets and computational speed. {\tt TLS} analyses the entire, unbinned phase-folded light curve. We compensated for the higher computational load by (i.) using algorithms such as ``Mergesort'' (for the trial orbital phases) and by (ii.) restricting the trial transit durations to a smaller range that encompasses all known planets, and using stellar density priors where available. A typical K2 light curve, including 80\,d of observations at a cadence of 30\,min, can be searched with {\tt TLS} in $\sim10\,$s real time on a standard laptop computer, as fast as the widely used box least squares ({\tt BLS}) algorithm. We perform a transit injection-retrieval experiment  of Earth-sized planets around sun-like stars using synthetic light curves with 110\,ppm white noise per 30\,min cadence, corresponding to a photometrically quiet $K_P=12$ star observed with Kepler. We  determine the SDE thresholds for both {\tt BLS} and {\tt TLS} to reach a false positive rate of 1\,\% to be SDE~=~7 in both cases. The resulting true positive (or recovery) rates are $\sim\,93$\,\% for {\tt TLS} and $\sim~76$\,\% for {\tt BLS}, implying more reliable detections with {\tt TLS}. We also test {\tt TLS} with the K2 light curve of the TRAPPIST-1 system and find six of seven Earth-sized planets using an iterative search for increasingly lower signal detection efficiency, the phase-folded transit of the seventh planet being affected by a stellar flare. {\tt TLS} is more reliable than {\tt BLS} in finding any kind of transiting planet but it is particularly suited for the detection of small planets in long time series from Kepler, TESS, and PLATO. We make our {\tt python} implementation of {\tt TLS} publicly available.}

\maketitle

\section{Introduction}
\label{sec:introduction}

Since the first discovery of an extrasolar planetary transit across the disk of a distant star \citep{2000ApJ...529L..45C}, exoplanet surveys have expanded greatly in numbers and volume. Ground-based transit searches such as HATNet \citep{2004PASP..116..266B}, WASP \citep{2006PASP..118.1407P}, KELT \citep{2007PASP..119..923P}, and CHESPA \citep{2018arXiv180901789Z}, and space-based search campaigns like CoRoT \citep{2009A&A...506..411A}, Kepler \citep{2010Sci...327..977B}, K2 \citep{2014PASP..126..398H}, and TESS \citep{2014SPIE.9143E..20R} produced vast data sets. These encompass hundreds of thousands of stars, cadences of seconds or minutes, and data sets that span several years. The PLATO space mission, with an expected launch in 2026 and a nominal six year duty cycle, will shadow these surveys by observing up to a million relatively bright stars with cadences of 25\,s or 10\,min \citep{2014ExA....38..249R}. These modern exoplanet transit searches require fast, sensitive, and reliable algorithms to detect the expected but unknown transit signals.

The box least squares ({\tt BLS}) algorithm \citep{2002A&A...391..369K,2016ascl.soft07008K} has become the standard tool for exoplanet transit searches in large data sets. It approximates the transit light curve as a (negative) boxcar function with a normalized average out-of-transit flux of zero and a fixed depth during the transit. This approach is key to its computational speed and allows for reliable detections of high to medium signal-to-noise ratio (S/N) signals, such as large (Jupiter-sized) and moderately large (Neptune-sized) planets around sun-like stars in most surveys.

The {\tt BLS} detection efficiency for low-S/N signals from Earth-sized planets around sun-like stars, however, is significantly smaller because the transit depths are comparable to the level of instrumental and stellar noise. Moreover, {\tt BLS} introduces a systematic noise component that comes from the mathematical concept of the boxcar function. This binary model of a fixed out-of-transit and a fixed in-transit flux is equivalent to the neglect of the stellar limb darkening and of the planetary ingress and egress in the light curve. This box-shape approximation introduces an extra noise component in the test statistic that dilutes low-S/N signals. Here we present an improved transit search algorithm that attempts to minimize this systematic noise component in the search statistic.

The {\tt BLS} algorithm has been analyzed and optimized in depth, for example in terms of the optimal frequency sampling, optimal phase sampling, and various other parameters \citep{2014A&A...561A.138O}. {\tt BLS} has been extended to variable intervals between successive transits \citep{2013ApJ...765..132C}, to work with non-Gaussian errors \citep{2014IAUS..293..410B}, to improve speed at the cost of sensitivity \citep{2008A&A...492..617R}, and to refine the detected transit parameters \citep{2006MNRAS.373..799C,2016A&C....17....1H}. Further adaptions were made for the application to circumbinary planets \citep{2008MNRAS.387.1597O}.

Alternatively, the ``matched filter'' algorithm is similar to {\tt BLS} in modeling the transit as a boxcar, but it uses a different test statistic \citep{1996Icar..119..244J,2007ASPC..366..145B}. Phase dispersion minimization \citep{1978ApJ...224..953S} has been shown to be inferior to {\tt BLS} for transit detection \citep{2002A&A...391..369K}. Analysis of variance \citep[AoV,][]{2006MNRAS.365..165S} also uses a box-shaped transit model (which the authors refer to as top-hat), and has been demonstrated to have a lower detection efficiency than {\tt BLS} in WASP data \citep{2012A&A...548A..48E}. Bayesian algorithms can search for any signal form \citep{2000ApJ...535..338D,2001A&A...365..330D}, but are not widely used. For example, the Gregory-Loredo method for Bayesian periodic signal detection uses step-functions (boxes with multiple steps) \citep{2002A&A...395..625A,2004MNRAS.350..331A}. Wavelet-based algorithms \citep{2007A&A...467.1345R} are of similar detection efficiency and have been widely used for automated analyses of CoRoT \citep{2009IAUS..253..374R}, Kepler \citep{2010ApJ...713L..87J}, and TESS \citep{2016SPIE.9913E..3EJ} data. Polynomials have also been suggested to approximate transit shapes more adequately than boxes \citep{2012A&A...548A..44C} and this idea has been used \citep{2016AJ....151..171J,2018AJ....156...78L}, although without a comparison to {\tt BLS} in terms of detection efficiency and computational effort.

Comparisons of different algorithms showed that {\tt BLS} is the best of all known algorithms for weak signals \citep{2003A&A...403..329T,2003A&A...408L...5T}, but ``no detector is clearly superior for all transit signal energies'', which has been verified by empirical tests of the methods \citep{2005A&A...437..355M}.

New techniques have now arrived with the advent of artificial intelligence. Deep learning algorithms are usually trained with a series of transit shapes \citep{2018MNRAS.474..478P,2018AJ....155..147Z,2018MNRAS.478.4225A}. Random-Forest methods detect 7.5\,\% more planets than classical {\tt BLS} for low S/N transits \citep{2016MNRAS.455..626M} because (many different) real transit shapes are used instead of a box. Disadvantages of these methods include substantial computational requirements, high implementation complexity, and a difficulty in understanding the origin of the results due to the many abstraction layers.

Here we present a new transit search algorithm that is easy to use, publicly available,\footnote{\href{http://github.com/hippke/tls}{http://github.com/hippke/tls}} and has a detection statistic that is generally more sensitive than that of {\tt BLS}. Most important, it is optimized to find small planets in large data sets. The algorithm assumes a realistic transit shape with ingress and egress and stellar limb darkening \citep[as per][]{2002ApJ...580L.171M} using a predefined parameterization that we optimized based on all previous exoplanet transit detections. The resulting increase in the detection significance of the algorithm by 5--10\,\% comes at the toll of larger computational demands. Given the tremendous growth of available CPU power in the past 60 years \citep{Moore1965} and in particular since the publication of the {\tt BLS} algorithm in 2002, however, we argue that CPU margins are not as crucial to the detection of small planets as is the significance of the test statistic. That said, we have nevertheless optimized the algorithm for computational speed as far as possible.

Our algorithm is particularly suited for the detection of Earth-sized planets with Kepler/K2, TESS, or with the future big data sets from the PLATO mission. In fact, the improvements of our new algorithm are most substantial for small planets with few transits, which is a common characteristic of Earth-sized planets in the habitable zones around sun-like stars.

\section{Methods}
\label{sec:methods}

\begin{figure}[ht]
\includegraphics[width=1.\linewidth]{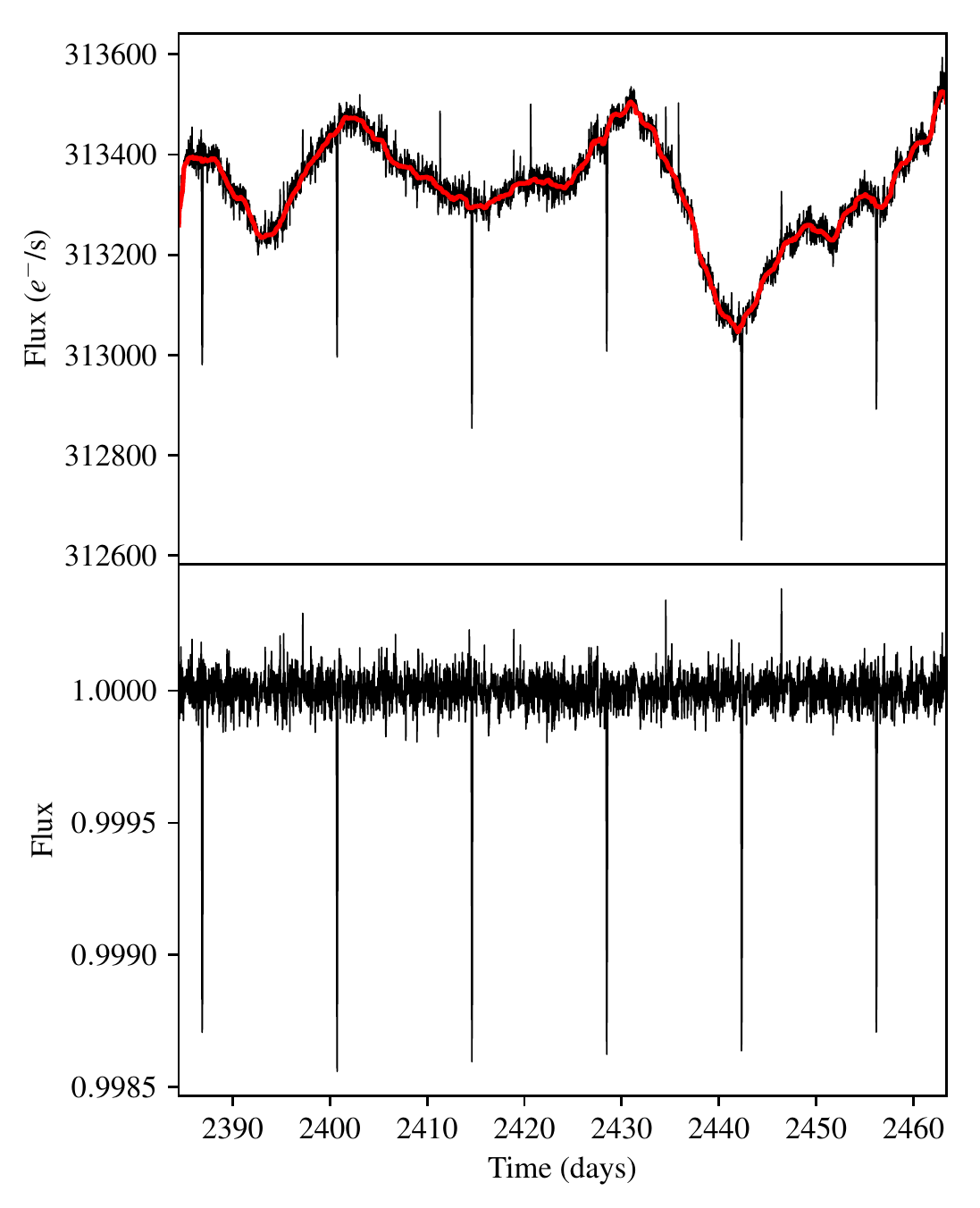}
\caption{\label{fig:k2-110b_lc}{\it Top:} K2 long cadence light curve of the star K2-110, which exhibits transits of a mini-Neptune-sized planet, K2-110\,b. The black line shows the light curve that has been corrected for instrumental effects with EVEREST and the red line shows our running median of 51 data points. {\it Bottom:} EVEREST light curve divided by the running median.}
\end{figure}

\begin{figure*}[t!]
\includegraphics[width=.5\linewidth]{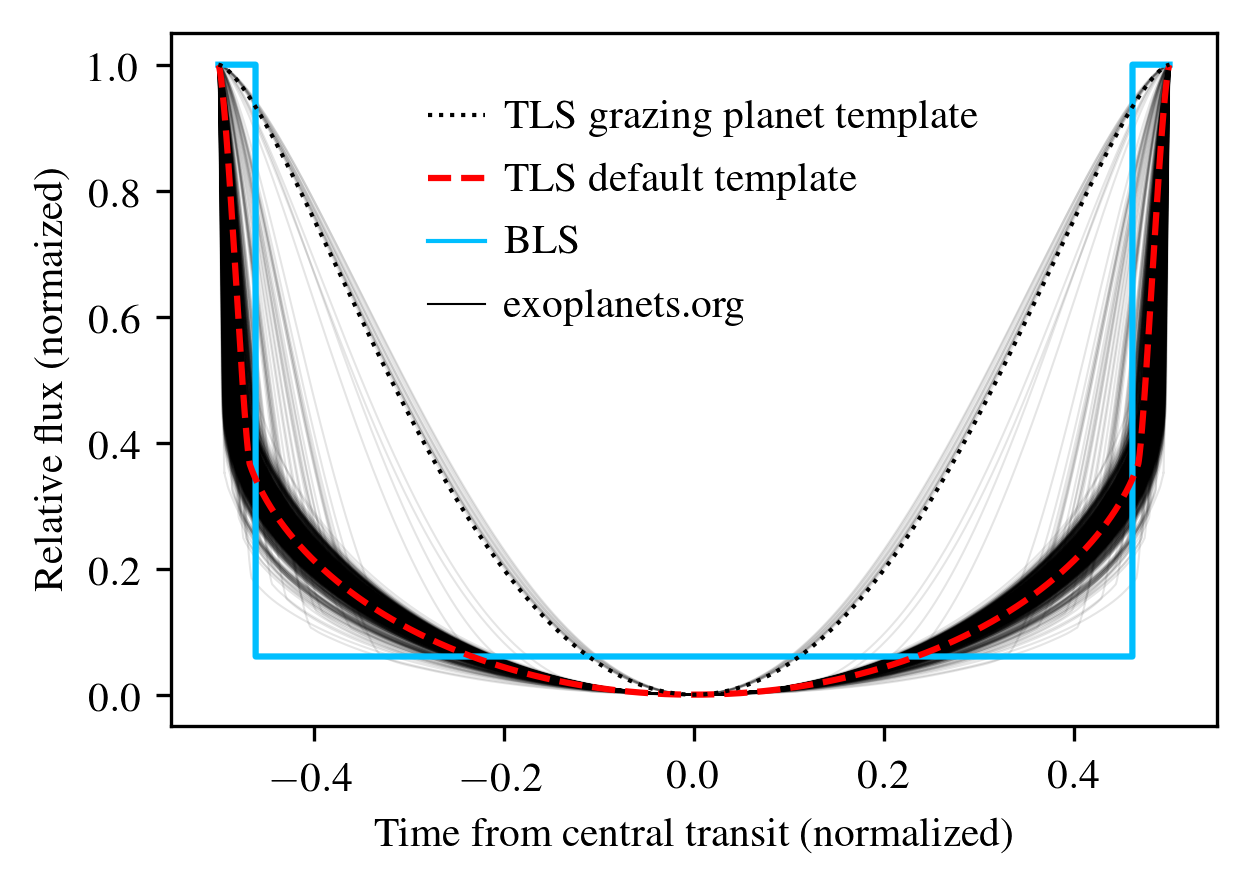}
\includegraphics[width=.5\linewidth]{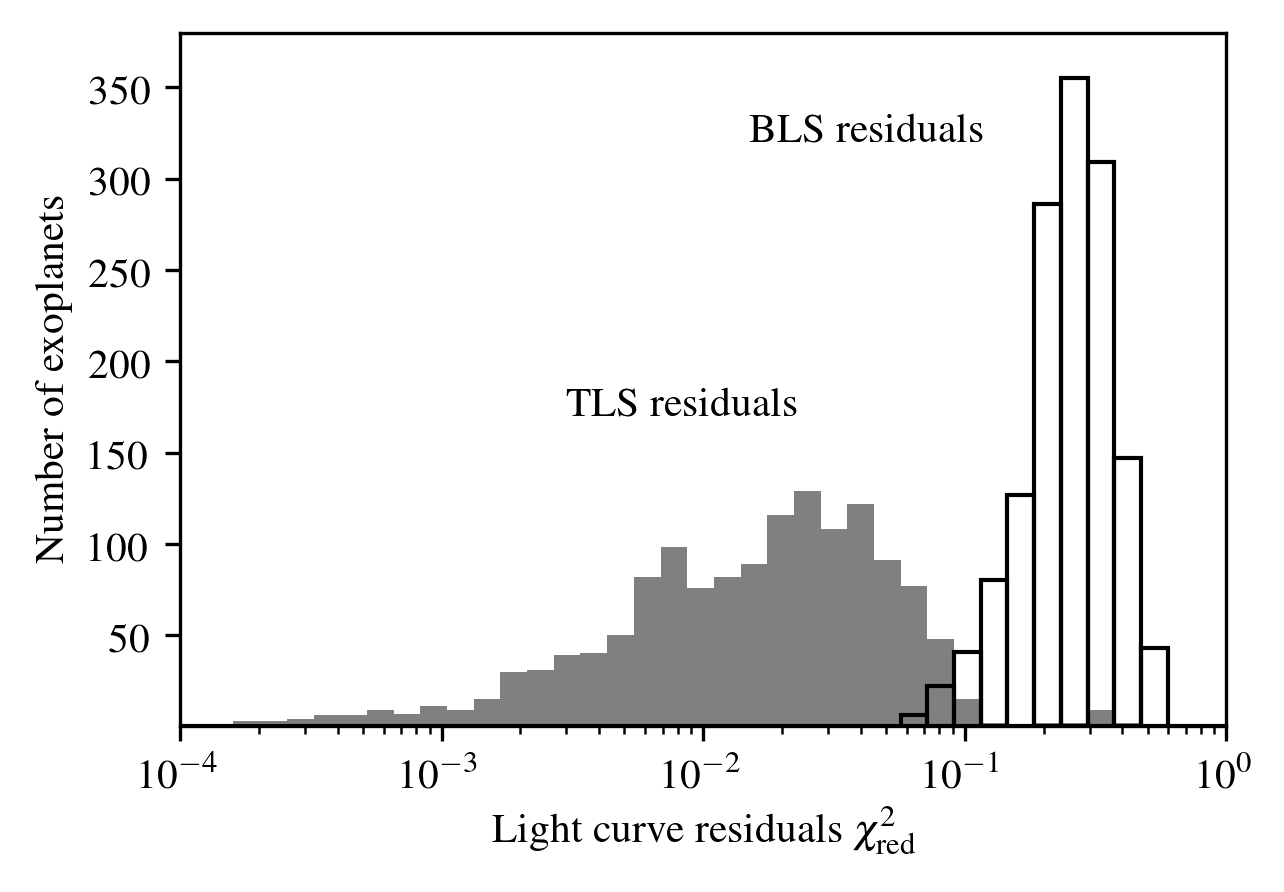}
\caption{\label{fig:transit_all_planets}{\it Left}: Transit light curves for all 2346 transiting Kepler planets from the Exoplanet Orbit Database (as of 1 November 2018) with $R_{\rm p}/R_{\rm s}<0.05\,{\sim}\,5.4\,R_{\oplus}/R_{\odot}$ (black lines, one per planet). The default template for {\tt TLS} is shown with a red dashed line and the optional grazing planet template is shown with a black dotted line. The best-fitting box is shown as a light blue solid line for comparison. {\it Right}: The gray histogram shows the reduced $\chi^2$ residuals between the {\tt TLS} default (median) transit template and the real transit light curves in the left panel. The open histogram shows the reduced $\chi^2$ residuals between the box and the real transit light curves. The separation between the two histograms confirms that the {\tt TLS} default transit template is a substantially better match to the observations than a box in $\gtrsim\,99$\,\% of the cases.}
\end{figure*}

We illustrate the methodology of {\tt TLS} using the K2 light curve of the metal-poor K3 dwarf star K2-110, which hosts a transiting massive mini-Neptune (K2-110\,b, EPIC\,212521166b) in a 13.86\,d orbit \citep{2017A&A...604A..19O}. \citet{2016ApJS..224....2H} estimate a stellar effective temperature of $T_{\rm eff}=4628\,$K, a surface gravity of $\log{(g)}=4.6$, a stellar mass of $M_{\rm s}=0.752\,M_{\odot}$, and a stellar radius of $R_{\rm s}=0.7\,R_{\odot}$.

In the top panel of Fig.~\ref{fig:k2-110b_lc} we show the K2 light curve after correction for instrumental effects with EVEREST \citep{2016AJ....152..100L} (black line) together with the running median of 51 exposures (red line). The lower panel of Fig.~\ref{fig:k2-110b_lc} shows the light curve after division by the running median. This is the data used throughout this section and we note that this pre-processing or detrending of the light curve is not part of {\tt TLS}.

Readers interested in detrending techniques are referred to the \citet{1964AnaCh..36.1627S} filter \citep[used by][]{2011ApJS..197....6G},
the median filter \citep{2013A&A...553A..30T},
polynomial filters \citep{2012ApJ...749...15G,2018A&A...617A..49R}, the Cosine Filtering with Autocorrelation Minimization \citep[CoFiAM,][]{2013ApJ...770..101K}, and Gaussian process \citep{2015MNRAS.447.2880A}.

\subsection{Transit model}

The key idea of {\tt TLS} is to search for transits using a transit-like search function rather than a box. Our first task then is to identify the transit light curve that is most representative of the known exoplanet transit light curves, assuming that future exoplanet detections will be done most efficiently with this particular function. We refer to this function as the default template for {\tt TLS} and describe its construction in the following. Although this decision of using a transit curve template to match previous detections might suggest that {\tt TLS} will inherit a detection bias and search mostly for planets like the ones we already know, we have verified that {\tt TLS} is better than {\tt BLS} in finding any kind of planet, in particular grazing transiters and very small planets.

In fact, we decide to optimize the {\tt TLS} template for the detection of small planets. We have verified that large planets, which produce deep transits, can also be found with this {\tt TLS} template with a higher signal detection efficiency than with a box. In turn, if we had chosen to optimize the template to find large planets, then small planets would be more likely to be missed. We impose an arbitrary limit of $R_{\rm p}/R_{\rm s}\,<\,0.05\,\sim\,5.4\,R_{\oplus}/R_{\odot}$ on the planet-to-star radius ratio and retrieve the orbital inclination ($i$), semimajor axis in units of stellar radii ($a/R_{\rm s}$), $R_{\rm p}/R_{\rm s}$, and the orbital period ($P$) for all transiting Kepler planets from the Exoplanet Orbit Database \citep[\href{http://exoplanets.org}{exoplanets.org};][]{2011PASP..123..412W}. We set the orbital eccentricity of each planet to zero and obtain the predicted limb darkening coefficients $c_1$ and $c_2$ of a quadratic limb darkening law for each host star, using the stellar effective temperature ($T_{\rm eff}$) and surface gravity ($\log(g)$), from the stellar model atmospheres of \citet{2012A&A...546A..14C,2013A&A...552A..16C} in the Kepler bandpass.

In the left panel of Fig.~\ref{fig:transit_all_planets} we plot the $2346$ resulting  model transit light curves, normalized to the transit depth and transit duration, using the \textsf{batman} implementation \citep{2015PASP..127.1161K,2015ascl.soft10002K} of the \cite{2002ApJ...580L.171M} analytic transit model with quadratic limb darkening (black lines, one for each planet). We then construct the {\tt TLS} default template transit curve from the median values of the above-mentioned input parameters to the analytic transit model for quadratic limb darkening (red dashed line). We emphasize that the {\tt TLS} template is not a fit to the observed normalized transit light curves in Fig.~\ref{fig:transit_all_planets} but a model light curve based on the median input parameters of all known transiting exoplanets. In fact, the user of {\tt TLS} is free to chose their own parameterization of a template transit light curve for their search. We also show our template for a grazing transiting planet (black dotted line) and the best fitting box.

In the right panel of Fig.~\ref{fig:transit_all_planets} we show the reduced $\chi^2$ residuals between the {\tt TLS} template and the real transit light curves (gray histogram) and the reduced $\chi^2$ residuals between the {\tt BLS} box and the real transit light curves (open histogram). The two histograms show a substantial offset with the {\tt TLS} template resulting in much smaller $\chi^2$ residuals than the box used for the {\tt BLS} algorithm. We note that the single outlier of the {\tt TLS} distribution at $\chi_{\rm red}^2\,\sim\,0.35$ belongs to KOI-7880, which is a grazing transit planet.

\subsection{Transit search statistic}
\label{sec:transit_search}

\begin{figure}
\includegraphics[width=\linewidth]{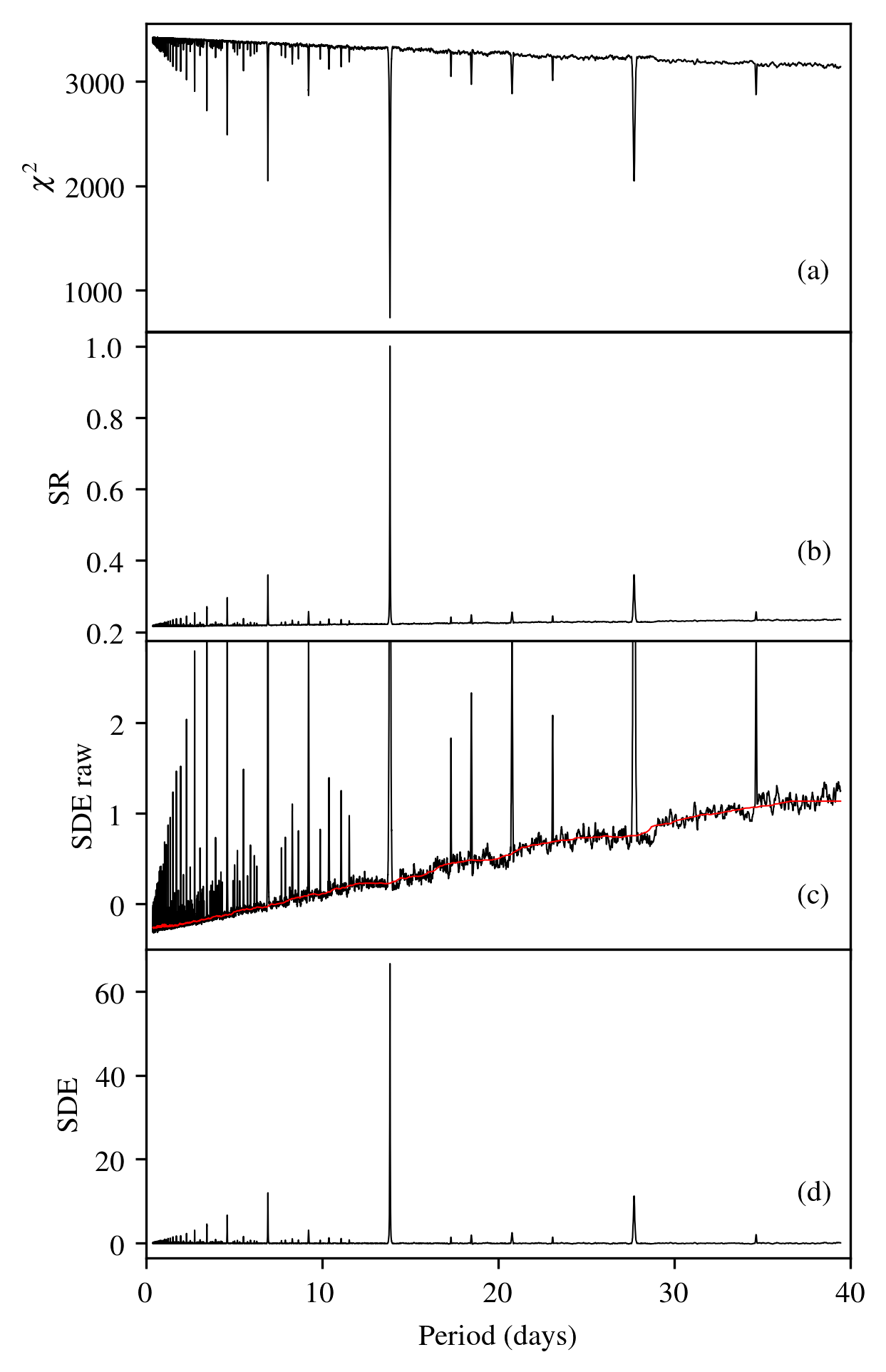}
\caption{\label{fig:spectra}(a): Distribution of $\chi^2$ (minimized over $t_0, d$) obtained by phase-folding the light curve of K2-110\,b over different trial periods. (b): Signal residue for the best fitting periods throughout the parameter space. (c): Raw signal detection efficiency (black line) and walking median (red line). (d): Signal detection efficiency used by {\tt TLS}. This is the result of the division of the raw SDE by its walking mean in panel (c).}
\end{figure}

The {\tt TLS} algorithm searches for periodic transit-shaped signals in time series of flux measurements. The algorithm operates by phase-folding the data over a range of trial periods ($P$), transit epochs ($t_0$), and and transit durations ($d$). It then calculates the $\chi^2$ statistic of the phase-folded light curve between the $N$ data points of the respective transit model $(y_i^{\rm m})$ and the observed values $(y_i^{\rm o})$ as per

\begin{equation}\label{eq:chisq}
\chi^2(P,t_0,d) =  \sum_{i=1}^N \  \frac{ (y_i^{\rm m}(P,t_0,d) - y_i^{\rm o})^2 }{\sigma_i^2} \ ,
\end{equation}

\noindent
where $\sigma_i^2$ are the standard deviations in the light curve. In Fig.~\ref{fig:spectra}(a), we show the spectrum of minimum $\chi^2$ as a function of $P$. In other words, for each trial period {\tt TLS} searches the minimum $\chi^2$ by testing all combinations of the ($t_0,d$) 2D hyperspace of the 3D parameter grid. {\tt TLS} uses the global $\chi^2$ minimum

\begin{equation}
\chi^2_{\rm min, glob} \equiv \min{\Big (} \ \chi^2(P,t_0,d) \ {\Big )} \equiv \chi^2(P', t_0', d')
\end{equation}

\noindent
at the location $(P', t_0', d')$ of our 3D parameter space for the normalization of the test statistic below. In Fig.~\ref{fig:spectra}(a), we locate $\chi^2_{\rm min, glob}$ at about 13.87\,d, corresponding to the published value of K2-110\,b by \citet{2017A&A...604A..19O}.

 {\tt TLS} uses a modified version of the test statistic originally implemented in the {\tt BLS} algorithm\footnote{In Appendix~\ref{sub:edge} we identify a glitch in a patch to the previously known {\tt BLS} edge effect that slightly affects the test statistic. This has been corrected for in {\tt TLS}.}, the signal detection efficiency \citep[SDE;][]{2000ApJ...542..257A}. The SDE has been widely demonstrated to yield useful results and it has become a standard metric in the exoplanet hunting community. Our implementation of {\tt TLS}, however, does not apply a binning of the phase-folded light curve to compute the signal residue (SR) between the model and the data. Our approach is computationally more expensive (quadratic in the number of data points, see Sect.~\ref{sec:costs}) but key to making {\tt TLS} more sensitive to the signals of small transiting planets.

 {\tt TLS} calculates the SR from the distribution of minimum $\chi^2$ as a function of $P$,

\begin{equation}\label{eq:SR}
{\rm SR}(P) = \frac{\displaystyle {\Bigg (} \frac{1}{\chi_{\rm min}^2(P)} {\Bigg )} }{\displaystyle {\Bigg (} \frac{1}{\chi^2_{\rm min, glob}} {\Bigg )}} = \frac{\chi^2_{\rm min, glob}}{\chi_{\rm min}^2(P)}
\end{equation}

\noindent
which necessarily results in the SR($P$) distribution to range between 0 and 1 (see Fig.~\ref{fig:spectra}b).

The SDE($P$) distribution is then obtained as per \citet{2002A&A...391..369K} using the arithmetic mean $\langle{\rm SR}(P)\rangle$, the standard deviation $\sigma({\rm SR}(P))$, and the peak value SR$_{\rm peak}$ of ${\rm SR}(P)$ via

\begin{equation}
{\rm SDE}(P) = \frac{{\rm SR_{\rm peak}} - \langle{\rm SR}(P)\rangle}{\sigma({\rm SR(P)})} \ .
\end{equation}

\noindent
With SR$_{\rm peak}\,=\,1$ by definition in Eq.~\eqref{eq:SR}, we have

\begin{equation}\label{eq:SDE}
{\rm SDE}(P) = \frac{1 - \langle{\rm SR}(P)\rangle}{\sigma({\rm SR}(P))} \ .
\end{equation}

\noindent
An SDE value of $x$ for any given $P$ means that the statistical significance of this period is $x\,\sigma$ compared to the mean significance of all other periods. We refer to the resulting SDE($P$) distribution in Fig.~\ref{fig:spectra}(c) as the raw SDE. The final step in our construction of the transit search statistic is in the removal of the systematic noise component that is inherent to the SDE distribution as explained by \citet{2014A&A...561A.138O}. We follow this author in removing this trend with a walking median filter through the SDE($P$) periodogram, the result of which is shown in Fig.~\ref{fig:spectra}(d) for K2-110. The transit signal of Kepler-110\,b can be found at 13.8662 with an SDE value of 66.7. We detetermine a conservative error estimate via the half width at half maximum of the SDE peak, which is 0.0122\,d. Hence, {\tt TLS} determines the orbital period of K2-110\,b as $P~=~13.8662\,(\pm\,0.0122)$\,d.

Empirical thresholds for a transit detection have been proposed to range from ${\rm SDE}>6$ \citep{2015ApJ...807...45D}, ${\rm SDE}>6.5$ \citep{2018AJ....156...78L}, ${\rm SDE}>7$ \citep{2012ApJ...761..123S}, 6--8 as a function of period \citep{2016MNRAS.461.3399P,2016MNRAS.459.2408A}, and up to 10 \citep{2018MNRAS.473L.131W}. Lower SDE thresholds mean better completeness but also higher false alarm rates.

\begin{figure}
\includegraphics[width=\linewidth]{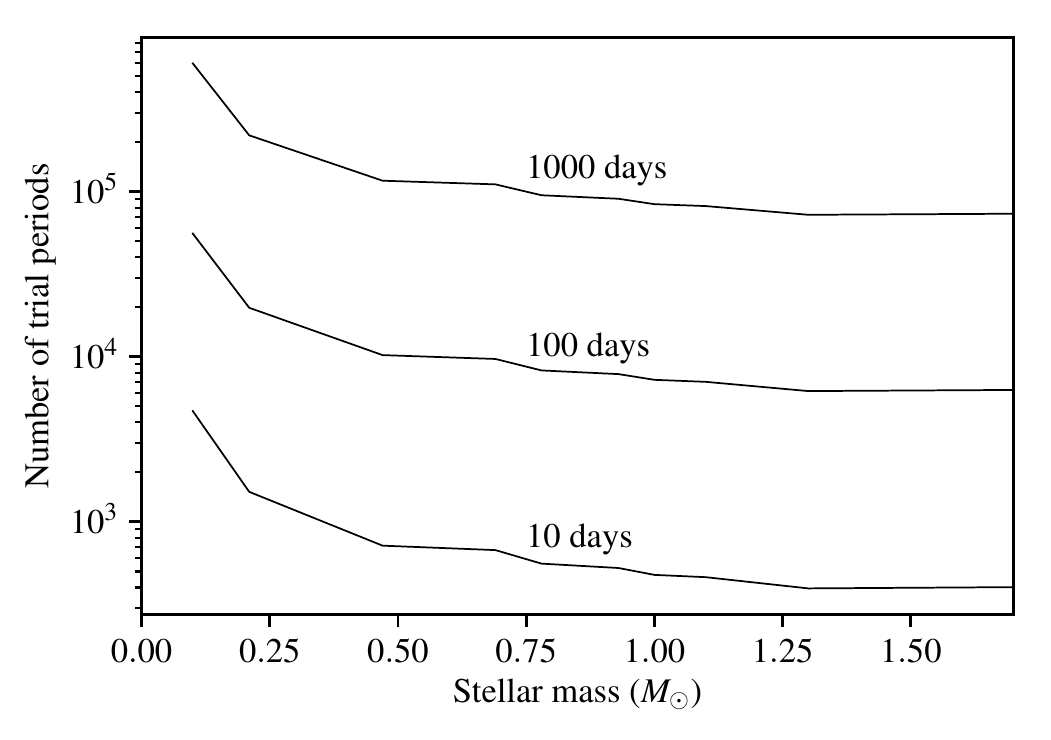}
\caption{\label{fig:sampling}Optimal number of trial periods ($N_{\rm freq,\,opt}$) as a function of stellar mass for three different time spans of a hypothetical stellar light curve.}
\end{figure}

\subsection{The {\tt TLS} parameter grid}

\subsubsection{Period sampling}
\label{sub:periods}

In the search for sine-like signals, for example using Fourier transforms, a uniform sampling of the trial frequencies is usually quite efficient. A uniform sampling of the orbital frequencies has also been suggested for {\tt BLS} \citep{2002A&A...391..369K}. As shown by \citet{2014A&A...561A.138O}, however, this sampling of the orbital frequency tends to be insensitive to either short- or long-period planets, but it is always computationally inefficient. \citet{2014A&A...561A.138O} derived the optimal number of test frequencies as

\begin{equation}
N_{\rm freq,\,opt} = \left( f_{\rm max}^{1/3} - f_{\rm min}^{1/3} + \frac{A}{3} \right) \frac{3}{A}
\end{equation}

\noindent
with

\begin{equation}
A=
\frac{(2\pi)^{2/3}}{\pi }
\frac{R_{\rm s}}{(GM_{\rm s})^{1/3}}
\frac{1}{S \times OS} \ ,
\end{equation}

\noindent
where $G$ is the gravitational constant, $S$ is the time span of the data set, and $OS$ is the oversampling parameter to be chosen between 2 and 5 to ensure that the SDE peak is not missed between trial frequencies (or periods). The minimum and maximum trial orbital frequencies can be found at $f_{\rm min}=2/S$ (or $f_{\rm min}=3/S$ if three transits are required) and at the most short-period (high-frequency) circumstellar orbit, the Roche limit, $f_{\rm max}= \sqrt{GM_{\rm s}/(3R_{\rm s})^3} / (2\pi)$. Strictly speaking, the Roche limit depends on the density ($\rho_{\rm p}$) of the planet, and the term $3\,R_{\rm s}$ for our expression of $f_{\rm max}$ assumes the most pessimistic case of an extremely low-density fluid-like planet with $\rho_{\rm p}~=~1\,{\rm g\,cm}^{-3}$, which can be compared to Jupiter's mean density of $1.33\,{\rm g\,cm}^{-3}$. Our {\tt TLS} implementation generates an array of evenly spaced orbital frequencies with $N_{\rm freq,opt}$ constant steps between $f_{\rm min}$ and $f_{\rm max}$ and then computes the (non-uniform) trial orbital periods as the inverse of this frequency grid.

Since computational speed is a key concern for us, we illustrate the resulting number of trial periods in Fig.~\ref{fig:sampling}, using three different time spans $S$ of a hypothetical light curve for different stellar masses (and radii, assuming main-sequence mass-radius relation). We find that an extension of the light curve by a certain factor -- here ten between the three example curves -- increases the number of trial periods by the same factor. This plot warns us of the large number of trial periods that need to be examined for planets around low-mass stars, with $N_{\rm freq, opt}$ reaching values of up to almost one million for a light curve with 1000\,d of continuous observations of a very-low-mass star. This feature is inherent to both {\tt TLS} and  {\tt BLS}.

\subsubsection{Transit depth}
{\tt TLS} measures the mean flux of the in-transit data points under consideration. It then calculates the corresponding maximum transit depth $\delta$ and the resulting planet-to-star radius ratio under the assumption of zero transit impact parameter using the analytic solutions for common stellar limb darkening laws found by \citet{Heller2018_LD}. With this calculation, the signal shape is scaled and compared to the data points.

{\tt} TLS can also be used to search for user-defined signal shapes (for example, flares), either with positive or negative flux. If an analytical scaling option is not available, {\tt TLS} can perform a numerical iterative fit using an initial guess based on the mean of the in-signal data, ${\langle}y_{i,{\rm in}}^o{\rangle}$. The $\delta$ range to be tested with {\tt TLS} is bracketed between ${\langle}y_{i,{\rm in}}^o{\rangle}/10$ and $10{\times}{\langle}y_{i,{\rm in}}^o{\rangle}$. {\tt TLS} uses an iterative ternary algorithm \citep{knuth1998art} to tighten the interval in steps of $1/3$ until the upper and lower limits differ by $<X\,$\% in signal depth (or height), where $X$ is a user-defined threshold.

\subsubsection{Transit duration}
\label{sub:transit_duration}

\begin{figure}
\includegraphics[width=\linewidth]{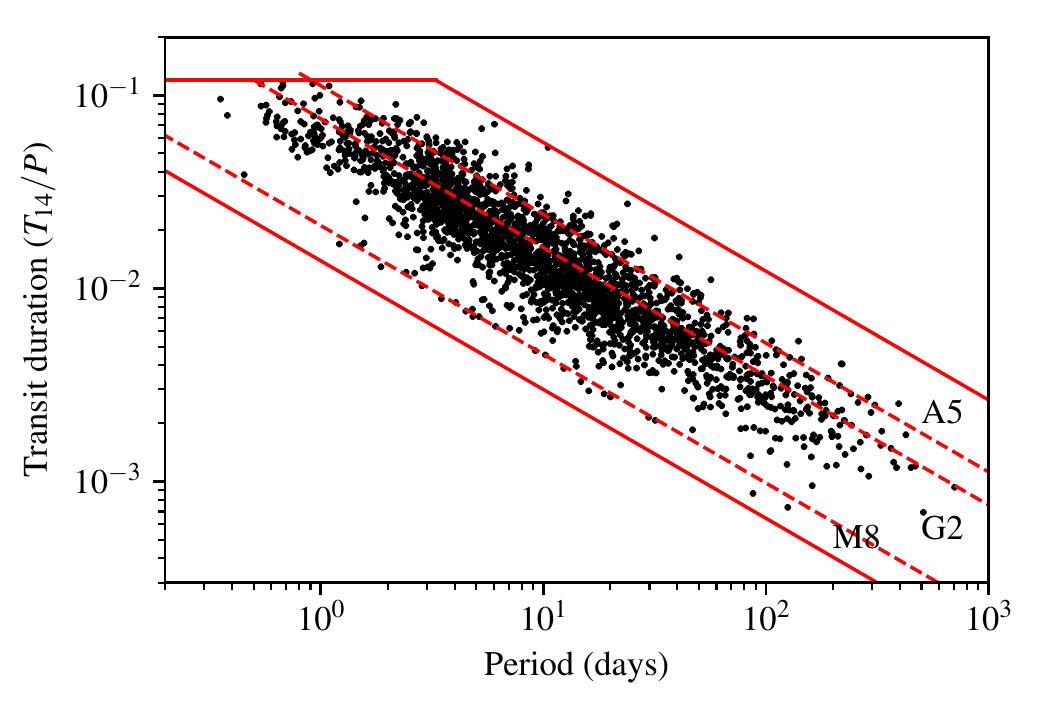}
\caption{\label{fig:per_t14}Transiting planets from the Exoplanet Orbit Database in the $T_{\rm 14}/P$-$P$ diagram. {\tt BLS} implementations typically search a linearly spaced uniform grid, or roughly the entire diagram. However, more than half of this search space is not populated with planets. The default parameterization of {\tt TLS} only searches inside the area embraced by the solid lines, which are defined in Eq.~\eqref{eq:T14max_2}. {\tt TLS} users can nevertheless redefine their own cuts of the $T_{\rm 14}/P$-$P$ diagram and search for planets with hitherto unknown properties in this diagram.}
\end{figure}

Transit searches using the {\tt BLS} algorithm usually operate with trial transit durations\footnote{We follow the common nomenclature to indicate the time interval between the first and fourth contact of the stellar and planetary silhouettes as $T_{14}$ \citep{2003ApJ...585.1038S}.} $T_{14}$ that span 0.00125~--~0.07 \citep{2013ApJ...770...69P}, 0.01~--~0.1 \citep{2012MNRAS.424.3101G}, or 0.001~--~0.2 \citep{2014ApJ...787...47S,2016MNRAS.459.2408A} times the orbital period. More than half of the corresponding $T_{14}(P)$ diagram, however, is not populated with exoplanet discoveries in these regions of the parameter space (see Fig.~\ref{fig:per_t14}). We explain this absence of transiting planets using geometrical constraints and Kepler's third law.

For example, there are no planets known with $P=10\,$d and $T_{\rm 14}/P<5\times10^{-3}$. From an astrophysical perspective, only extremely eccentric planets could have such a short transit duration -- apparently a very rare, or even non-existent kind of exoplanet. It thus appears reasonable to us to restrict the computational effort to the physically plausible regions of the $T_{\rm 14}/P$ diagram. We also conclude from Fig.~\ref{fig:per_t14} that the transit duration search grid shall be linear in log-space.

For circular orbits, the maximum transit duration is $T_{14,{\rm max}}~=~2(R_{\rm s}+R_{\rm p})/v_{\rm p}$, where $v_{\rm p}$ is the planet's average orbital velocity during the transit. Shorter transit durations are possible if the planetary transit path is not across the stellar diameter. We then have

\begin{equation}\label{eq:T14max_1}
T_{14,{\rm max}} = \frac{2\,(R_{\rm s}+R_{\rm p})}{v_{\rm p}} = \frac{2\,(R_{\rm s}+R_{\rm p})}{2\pi a / P}
\end{equation}

\noindent
In the limit of the star being much more massive than the planet, Kepler's third law becomes

\begin{align}\label{eq:Kepler3law}\nonumber
\left(\frac{2\pi}{P}\right)^2 \ a^3 \approx \ & G M_{\rm s} \\
\Leftrightarrow \hspace{1.8cm} a \approx \ & \left( G M_{\rm s} \left( \frac{P}{2\pi} \right)^2 \right)^{1/3} \ .
\end{align}

\noindent
We insert Eq.~\eqref{eq:Kepler3law} into Eq.~\eqref{eq:T14max_1} and obtain

\begin{align} \label{eq:T14max_2}\nonumber
T_{14,{\rm max}} = \ & \frac{(R_{\rm s}+R_{\rm p}) \, P}{\pi} \left( \frac{1}{G M_{\rm s}} \left( \frac{2\pi}{P} \right)^2 \right)^{1/3} \\
= \ & (R_{\rm s}+R_{\rm p}) \left( \frac{4P}{\pi G M_{\rm s}} \right)^{1/3} \ .
\end{align}

In Fig.~\ref{fig:per_t14} we plot Eq.~\eqref{eq:T14max_2} for $R_{\rm p}~=~2\,R_{\rm J}$ ($R_{\rm J}$ being Jupiter's radius) orbiting an A5 star ($M_{\rm s}=2.1\,M_\odot, R_{\rm s}=1.7\,R_\odot$) to maximize the effect of a very large planet on the transit duration. To embrace the physically plausible search space, we also show a main-sequence M8 red dwarf star ($M_{\rm s}=0.1\,M_\odot, R_{\rm s}=0.13\,R_\odot$) and a sun-like star, both with a small planet ($R_{\rm p}~=~\,R_{\rm \oplus}$). We see that a significant amount of planets is actually located above the uppermost of these lines, which can be attributed to bloated super-Jovian planets and/or to planets transiting slightly evolved stars, for example. The absence of planets in the lower left part of the diagram could be partly astrophysical in nature and interpreted as a trace of planet formation and evolution (for example, the absence of ultra-short period planets around M dwarfs). That said, planets can naturally have transit durations that are arbitrarily shorter than $T_{14,\rm{max}}$, for example, on eccentric or inclined orbits. The empty space can nevertheless be explained with a detection bias against planets with transit durations of just a few minutes, for example, $\sim\,15$\,min for $P=1$\,d and $T_{14}/P=10^{-2}$.

In order to compensate for planets transiting evolved stars as well as for planets on eccentric orbits\footnote{In eccentric cases the average orbital velocity in-transit can be smaller or larger and the resulting transit duration can be larger or smaller than in the circular case.} and other astrophysical effects that are potentially hard to predict, we use Fig.~\ref{fig:per_t14} to derive empirical estimates for the maximum and minimum values of $d$ to be searched. We parameterize the upper limit via Eq.~\eqref{eq:T14max_2} and using $M_{\rm s}=1\,M_\odot$ and $R_{\rm s}=3.5\,R_\odot$, and we parameterize the lower limit of $T_{14}$ using $M_{\rm s}=1\,M_\odot$ and $R_{\rm s}=0.184\,R_\odot$ in Eq.~\eqref{eq:T14max_2}. We note that these two parameterizations do not correspond to any particular or even physically plausible main-sequence star. The motivation behind this parameterization of Eq.~\eqref{eq:T14max_2} is entirely empirical with the aim of embracing all known transiting exoplanets. Searches for planets around more exotic stars, such as white dwarfs, require other limits. Using our {\tt TLS} implementation, the user can conveniently set arbitrary limits of their choice.

Our default empirical limits for the transit durations to be searched with {\tt TLS} are shown with inclined solid lines in Fig.~\ref{fig:per_t14} and their parameterization was intentionally chosen to encompass all known transiting exoplanets. The horizontal cutoff at $T_{14}/P~=~1.12~{\times}~10^{-1}$ is a global threshold. This is the default region in the $T_{14}$-$(P)$ diagram to be tested with our implementation of the {\tt TLS} algorithm. That said, the user can define their own range of transit durations to be tested.

\section{Results}

\begin{figure*}
\includegraphics[width=.5\linewidth]{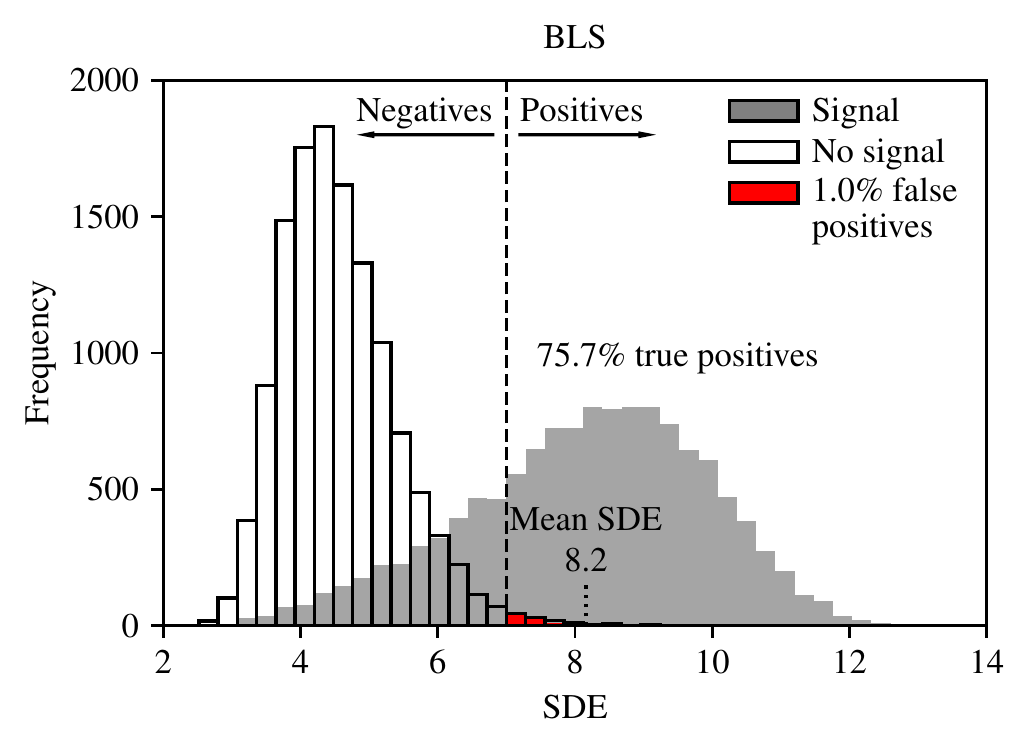}
\includegraphics[width=.5\linewidth]{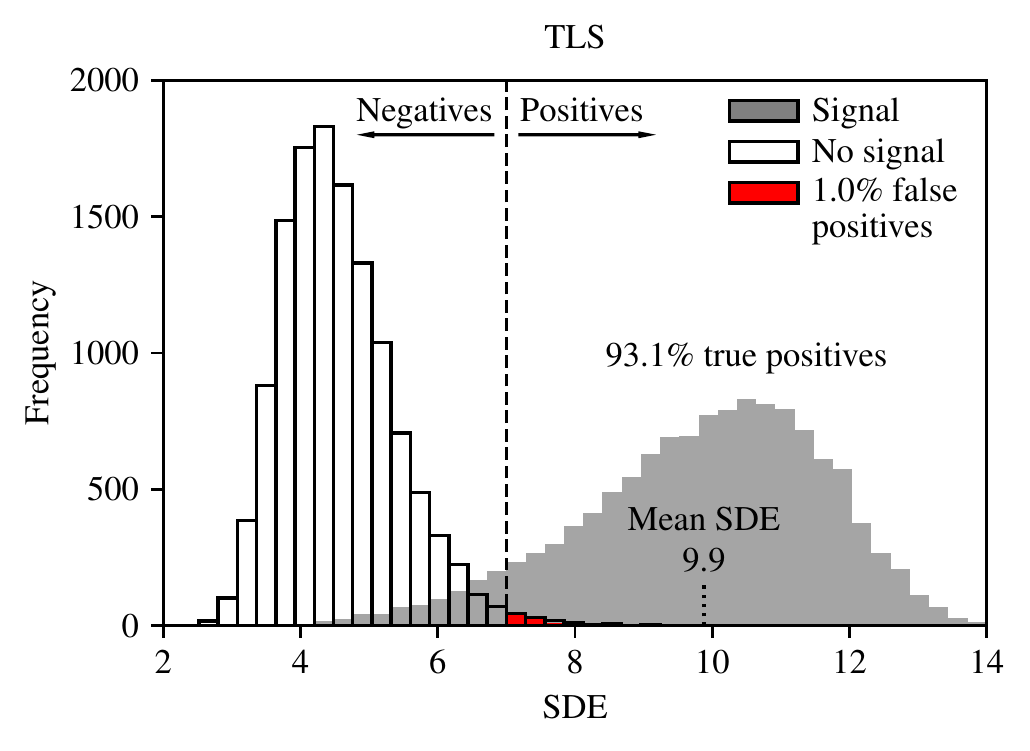}
\caption{\label{fig:histos_noise}Statistics of the signal detection efficiency for a transit injection-retrieval experiment of simulated light curves with white noise only. {\it Left}: box least squares algorithm. {\it Right}: transit least squares algorithm. Both panels show the results of $10\,000$ realizations of a 3\,yr light curve with white noise only (open histograms) and of the same amount of light curves with white noise and an Earth-like planetary transit around a G2V star (gray histograms). Outlined histograms relate to the SDE maximum value in a noise-only search. Gray histograms refer to the highest SDE value within 1\,\% of the period of the injected transit. The SDE thresholds at which the false positive rates are 1\,\% is found to be ${\rm SDE}_{\rm fp=1\,\%}=7$. At this SDE threshold, the recovery rate of the injected signals (the true positive rate) is 75.7\,\% for {\tt BLS} and 93.1\,\% for {\tt TLS}, while the fraction of missed signals (the false negative rate) is 24.3\,\% for {\tt BLS} and 6.9\,\% for {\tt TLS}.}
\end{figure*}

\subsection{{\tt TLS} and {\tt BLS} signal detection efficiency for white noise}
As a first test of the performance of {\tt TLS} in comparison to {\tt BLS} we generated synthetic light curves with white noise only. These light curves have a time span of $S~=~3$\,yr and a cadence of 30\,min with 110\,ppm noise (standard deviation) per cadence. This noise level is adapted to a best-case scenario for Kepler data, where the total noise over 6.5\,hr was found to be 30\,ppm for a $K_P=12$ star \citep{2015AJ....150..133G}. We note that the noise in the Kepler light curves does not have Gaussian properties due to instrumental and stellar trends (for example, from stellar rotation), all of which complicates transit detections in practice.

We injected three transits of an Earth-sized planet around a sun-like star with transit impact parameters randomly chosen in $b=[0,1]$ and with solar quadratic limb darkening as seen in the Kepler bandpass into a set of $10\,000$ pre-computed synthetic light curves with different white noise realizations. Then we conducted both a {\tt TLS} search and a standard {\tt BLS} search, in both of which we used our optimized period grid (Sect.~\ref{sub:periods}) for a fair comparison. A detection was counted as ``positive'' if the highest peak in the power spectrum was within 1\,\% of the injected transit period.

We determine the SDE thresholds for a false positive rate of 1\,\% to be ${\rm SDE}_{\rm fp=1\,\%}=7$ for {\tt BLS} and {\tt TLS}. Given these thresholds and the (forced) false positive rates of 1\,\%, {\tt TLS} recovers 93.1\,\% of the injected signals (the true positives) compared to 75.7\,\% for {\tt BLS}. While the SDE distribution of the white noise-only light curves is virtually identical for both {\tt BLS} and {\tt TLS}, the SDE distribution of the light curves with signal is shifted to higher SDE values for {\tt TLS}, with a mean of $\langle{\rm SDE }\rangle_{{\tt TLS},p}=9.9$ compared to a mean of $\langle{\rm SDE }\rangle_{{\tt BLS},p}=8.2$ for {\tt BLS}. In Fig.~\ref{fig:histos_noise}, we illustrate the results of our experiment.

\begin{figure}
\includegraphics[width=\linewidth]{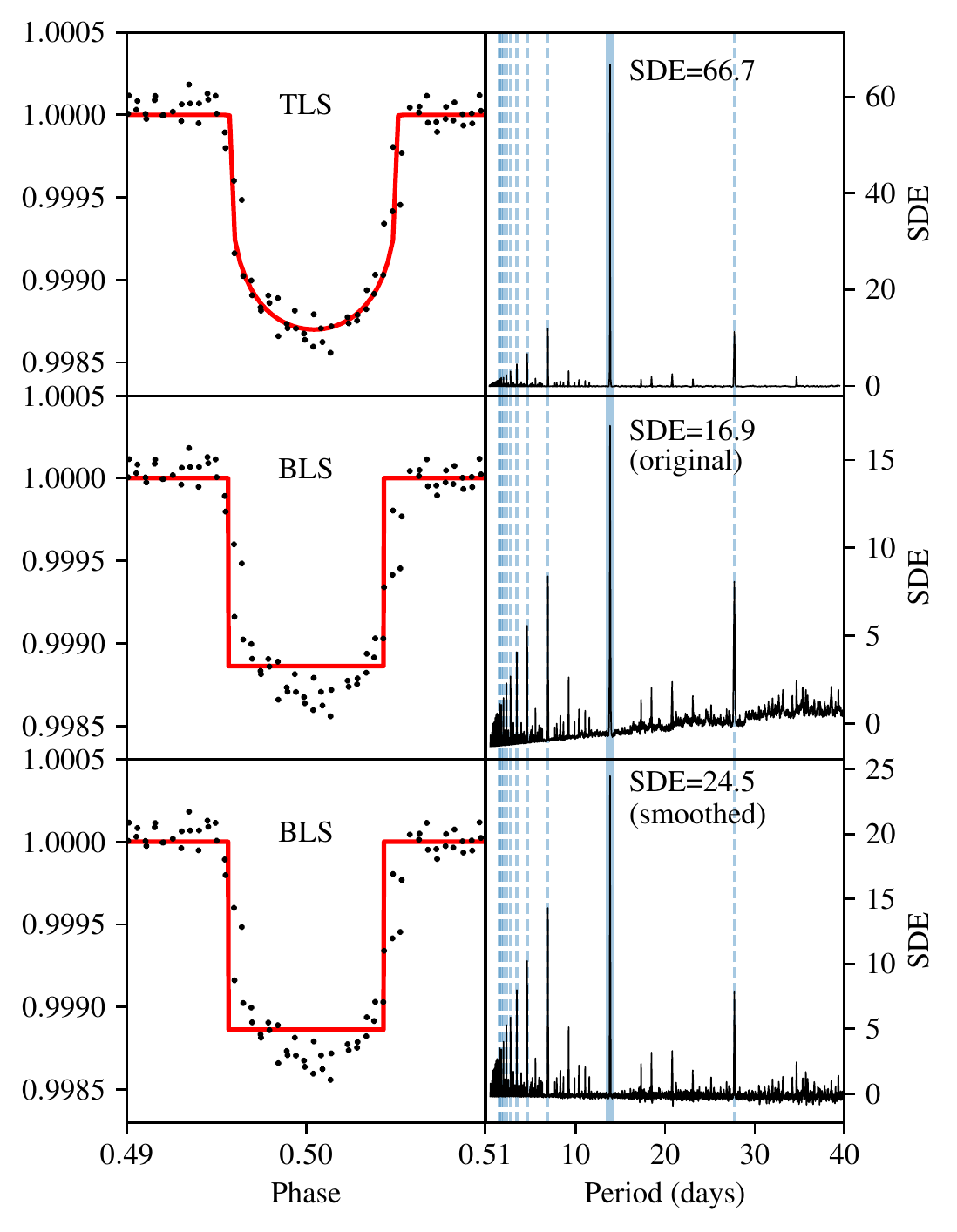}
\caption{\label{fig:k2-110b-comp}Phase-folded transits and peridograms of Kepler K2-110\,b with {\tt TLS} (top) and {\tt BLS} (middle, bottom) fitting for the same trial periods and durations. We note the boost in signal detection efficiency from 16.9 with the original {\tt BLS}, or 24.5 with the median-smoothed {\tt BLS} to 64.2 with {\tt TLS}. The vertical dashed blue lines denote the aliases of the period detected at the highest SDE value, respectively.}
\end{figure}

\begin{figure}
\centering
\includegraphics[width=1\linewidth]{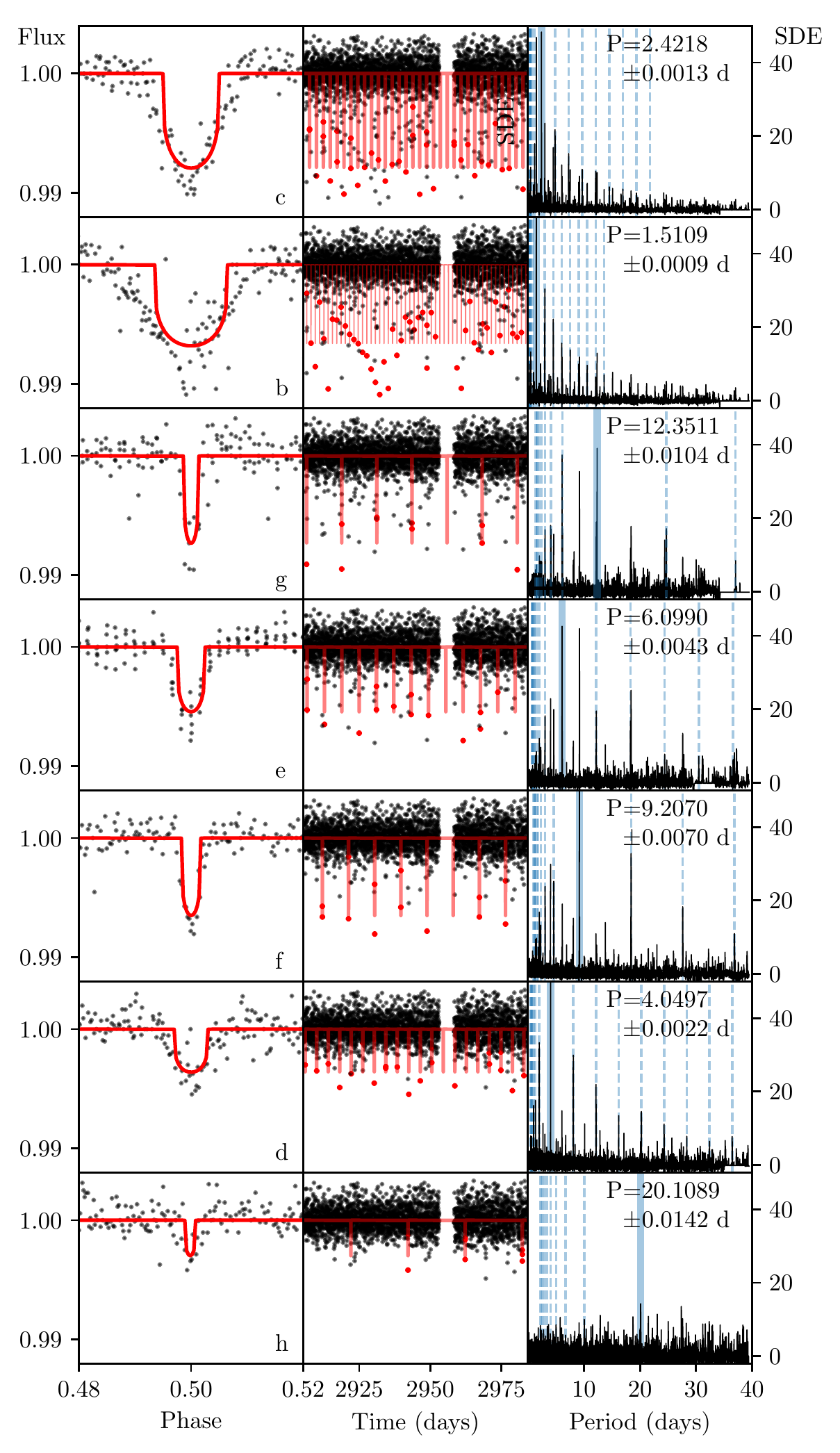}
\caption{\label{fig:trappist}Demonstration of the {\tt TLS} performance on the TRAPPIST-1 system. {\it Left}: Phase-folded transit light curve for the respective period and epoch at SDE maximum (black dots). The best-fit transit model (fitted for transit duration and depth) with quadratic stellar limb darkening is shown with a red solid line. Planet names are indicated in the lower right corner of each panel. Planets are sorted from top to bottom in the order of detection from an iterative {\tt TLS} search of the K2 light curve. Planet ``h'' (bottom panels) is a false positive and not related to the actual detection of TRAPPIST-1\,h (see Sect.~\ref{sec:TRAPPIST1}). {\it Center}: The entire K2 light curve of TRAPPIST-1 with the detected in-transit data points highlighted in red. Transits detected in previous iterations were masked. {\it Right}: SDE($P$) diagram for the light curve shown in the center.}
\end{figure}

\subsection{Comparison of {\tt TLS} and {\tt BLS} for K2-110\,b}
\label{sec:K2-110b}
We now compare the performance of {\tt TLS} and {\tt BLS} using a known planet with a high signal-to-noise ratio, K2-110\,b (EPIC~212521166), a massive mini-Neptune orbiting an old, metal-poor K3 dwarf star with a 13.9\,d period \citep{2017A&A...604A..19O}. We retrieve estimates for the stellar mass ($0.752\substack{+0.053\\-0.044}\,M_{\odot}$), radius ($0.7\substack{+0.048\\-0.045}\,R_{\odot}$), effective temperature ($T_{\rm eff}=4841\substack{+174\\-145}\,$\,K), and surface gravity ($\log{(g)}=4.628\substack{+0.025\\-0.03}$) from the Kepler K2 EPIC catalog \citep{2016ApJS..224....2H} using the automated {\tt catalog\_info} function of {\tt TLS}. {\tt catalog\_info} also retrieves the quadratic limb-darkening coefficients for the Kepler bandpass ($a=0.7010$, $b=0.0462$) via a cross-match of the \citet{2012A&A...546A..14C} tables based on $T_{\rm eff}$ and $\log{(g)}$. With these priors, {\tt TLS} creates an optimal period grid with an oversampling factor of five, which results in $21\,500$ trial periods between $P_{\rm min}=0.4\,$d and $P_{\rm max}=d/2=40\,$d.

We run {\tt BLS} and {\tt TLS} searches with the same period grid, oversampling, and duration constraints (Fig.~\ref{fig:k2-110b-comp}). {\tt TLS} delivers a much higher SDE (66.7) compared to {\tt BLS} (16.9, or 24.5 when median-smoothed). This is despite the fact that our priors are slightly different to the improved posteriors from \citet{2017A&A...604A..19O}, which suggest a hotter star ($T_{\rm eff}=5{,}050\pm50\,$K) with the same surface gravity, resulting in different quadratic limb darkening parameters ($a=0.5322$, $b=0.1787$).

\subsection{Recovery of the TRAPPIST-1 planets}
\label{sec:TRAPPIST1}
Moving on to real light curves of Earth-sized planets with time-correlated (red) noise components, we chose the K2 light curve of TRAPPIST-1 as a stress-test for {\tt TLS} to ensure its robustness. The system exhibits noise from instrumentals, stellar rotation, flares, and other sources, which can only imperfectly be removed using detrending. Signals of multiple planets occur with overlapping transits. Each planet produces its own set of harmonics and subharmonics in the power spectrum. To be considered robust, a detection algorithm must able to handle these difficulties and our attempt of recovering this previously reported series of transit signals in retrospect offers an exquisite case to simulate the {\tt TLS} search for Earth-sized planets.

The well-studied TRAPPIST-1 system exhibits transits of seven terrestrial planets \citep{2016Natur.533..221G,2017Natur.542..456G} in a resonant chain, where the orbital periods are near-ratios of small integers \citep{2017NatAs...1E.129L}. An automatic recovery of all planets is certainly difficult because of the low S/N of the individual transits resulting from the dim host star, the very small (Earth-sized) planets, transit timing variations, stellar flares, systematic trends from the stellar rotation of 3.3\,days, and overlapping multi-planet transits.

As before, we use the K2 EVEREST data spanning 79 days in campaign 12. We divide the data by a running median of 13 data points.

On the one hand, the length of the running median filter window must be larger than the transit duration to prevent the transit signal from being distorted by the median filter. On the other hand, the window length must be sufficiently short to remove stellar variability. For planets around TRAPPIST-1, the longest plausible transit duration at a period of 40 days is 1.6\,hrs (3 cadences). Our choice of 13 cadences is $\sim4$ times longer than the critical value. Longer median filter windows increase the residuals of the stellar noise significantly due to the strong variability of TRAPPIST-1. For comparison, in the case of K2-110b (Sect.~\ref{sec:K2-110b}), the amplitude of the stellar variability is much smaller and it occurs on timescales that are much longer than the transit duration. Therefore, we set the window length to 51 points ($\sim2$\,d), but a window length of 1\,d yields virtually the same result.

We then remove data points that deviate positively from the mean flux by more than $+3\,\sigma$ to eliminate bright flares. A detailed analysis by \citet{2018AJ....156..218D} carefully identifies transits affected by flares, incomplete transits, and multi-planet transits, which increases the quality of the in-transit data. Such a fine-tuned processing is beyond the scope of our analysis.

For our {\tt TLS} search, we use the default template and priors on the stellar mass ($0.089 \pm 0.006\,M_{\odot}$) and radius ($0.121 \pm 0.003\,R_{\odot}$) with limb darkening for the Kepler bandpass of an M8-star \citep{2018ApJ...853...30V}. The first run results in an ${\rm SDE}\sim45$ detection of a signal with a period of $P=2.4218\pm0.0013\,$d, which we identify as planet ``c'' (Fig.~\ref{fig:trappist}). Then we mask the in-transit data points of this signal using the {\tt TLS} convenience function \texttt{transit\_mask} and re-run {\tt TLS} iteratively. Each successive run results in the detection (and masking) of planets c--b--g--e--f--d. The order of detection is based on the signal-to-noise ratio of the stacked transits. While planet ``b'' nominally has the highest S/N, its transit shape differs significantly from the {\tt TLS} default template, making planet ``c'' have the highest SDE in the first {\tt TLS} run. The seventh and outermost planet ``h'' is not automatically detected by {\tt TLS}. We attribute this to the low number of transits (four) and to a flare that happened during the fourth transit, resulting in several in-transit data points showing unusually high flux. We show the highest SDE peak (caused by noise) of this last search in the bottom row of Fig.~\ref{fig:trappist} together with the best-fit transit shape, which is very noisy. This false positive signal illustrates the limits of automatic planet recovery, which apply to both {\tt TLS} and {\tt BLS}. We also verified that {\tt BLS} is not able to detect planet ``h'' using the same data processing.

\begin{figure}
\includegraphics[width=\linewidth]{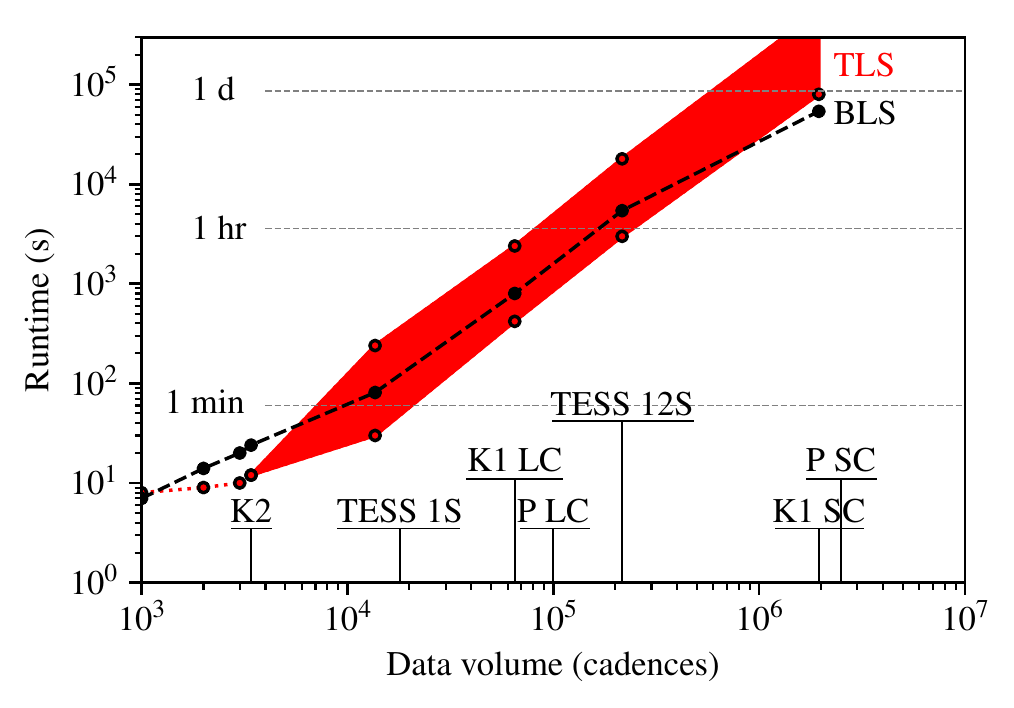}
\caption{\label{fig:speed}Algorithm run times for different missions, durations, and cadences. Kepler LC (30\,min) and SC (1\,min) are shown for 4.25\,yr worth of data, K2 assumes 80\,d of LC data. TESS is represented with one (1S) and twelve (12S) seasons at 2\,min cadence, respectively. PLATO light curves are considered over 2\,yr in both 25\,s short cadence (P SC) and 10\,min long cadence (P LC). The red area shows the full range of {\tt TLS} run times. The upper end assumes no priors on stellar density and fitting signals down to 10\,ppm. The lower end assumes typical priors from catalog data and a 100\,ppm threshold (or a 1\,\% threshold on phase sampling). In the latter case, run times are shorter than {\tt BLS} for all but the largest data sets. We note the slope of roughly two orders of magnitude of the run time per order of magnitude of cadences.}
\end{figure}

\subsection{Computational costs}
\label{sec:costs}

{\tt TLS} aims at maximizing sensitivity while our implementation of {\tt TLS} aims at maximizing computational speed at the same time. Since computing power has been continuously increasing for more than half a century now \citep{Moore1965} and since the whole point of {\tt TLS} is to offer unprecedented sensitivity, we prioritized the latter over computational efficiency whenever necessary (for example: no phase binning).

The computational effort per light curve is a complicated function of the stellar mass and stellar radius (see Fig.~\ref{fig:sampling}) and it depends linearly on the time span of the light curve ($S$). Another important factor is the number of trial epochs (or trial phases), which increases linearly with $S$ for a constant cadence. Both things combined, we find that the computational load increases quadratically with $S$, which itself is proportional to the number of data points for a fixed cadence.

The default {\tt TLS} configuration has a typical run time per Kepler K2 long cadence (K2 LC) light curve ($\sim4000$ data points, $S~=~80$\,d) of 10\,s, virtually identical to {\tt BLS} on the same Intel Core i7-7700K (Fig.~\ref{fig:speed}). We used the reference implementation of {\tt BLS} provided by {\tt Astropy~3.1} \citep{2013A&A...558A..33A,2018AJ....156..123A} in the C programming language, parallelized with the OpenMP interface\footnote{\url{http://docs.astropy.org/en/stable/api/astropy.stats.BoxLeastSquares.html}}. To compare K2 run times, we used the same number of trial transit durations (66) in both algorithms, and the same optimal grid of $\sim10000$ periods as determined by {\tt TLS}. We note that the optimal grid is not available by default in {\tt Astropy} and most other {\tt BLS} implementations, and only used by part of the community, resulting in three to five times longer run times for {\tt BLS} at the same level of sensitivity. {\tt TLS} run times are strongly dependent on the stellar density prior and the shallowest transit depth considered for fitting. A range of plausible values for optimistic and pessimistic cases is shown in Fig.~\ref{fig:speed} and explained in more detail in Appendix~\ref{sec:ap_speed}.

Our measurements are in agreement with ``one minute run time [for BLS] per processor core per K2 campaign star'' \citep{2016ApJS..222...14V}. Some years ago, {\tt BLS} performance was noted as ``10 minutes run time on a desktop workstation'' for a MEarth star with $1000$ data points \citep{2012AJ....144..145B}. {\tt BLS} run times of the \texttt{PyKE kepbls} routine have been reported as ``26 minutes (...) using a 3\,GHz Intel Core 2 Duo'' \citep{2012PASP..124..963K}.\footnote{\href{https://keplergo.arc.nasa.gov/ContributedSoftwareKepbls.shtml}{https://keplergo.arc.nasa.gov/ContributedSoftwareKepbls.shtml}} {\tt BLS} speedup factors of $25\,\times$ for K2 sized data are projected using optimal period sampling and optimal phase sampling \citep{2014A&A...561A.138O} for nominal {\tt BLS} sensitivity.

Longer data sets such as the long cadence (LC) light curves from the Kepler primary mission (K1) with $\sim60\,000$ data points and $S~=~4.25$\,yr are more demanding, as are K1 short cadence (K1 SC) data. PLATO long cadence (P LC) light curves will have a 10\,min sampling over 2\,yr (per field of view) and about $10^5$ cadences. PLATO short cadence (P SC) will be 25\,s and deliver $2.5\times10^6$ cadences over 2\,yr per light curve.

All quoted {\tt TLS} run times include the calculation of initial star-specific templates, which requires $\sim10\,$ms for the quadratic limb-darkening law. We explain these technicalities in more detail in the Appendix~\ref{sec:ap_speed}.

\subsection{Comparison to other transit detection algorithms}

\citet{2016MNRAS.455..626M} claim that their machine-learning method is 1000 times faster than {\tt BLS}, but do not include (or state) the run time required for the training part. The method is described to detect 8\,\% more planets compared to {\tt BLS}. However, no false/true positive/negative rates are given (as in our Figure~\ref{fig:histos_noise}), preventing a comparison to {\tt TLS}.

Training times for algorithms based on deep-learning to detect transits are of order several thousand CPU hours \citep{2018AJ....155...94S}. It remains unclear whether the training could be re-used in later searches. What is more, the authors describe a drop in model performance toward lower S/N transits, because few (real world) training candidates with low S/N were available.

Other studies find that random forest classifiers and convolutional neural networks produce a significant fraction of false-positives \citep{2018arXiv181107754S}. Depending on the threshold of a detection, it may also result in the outcome that {\tt BLS} has the smallest fraction of false negatives (missed detections), 5\,\% versus 11--14\,\% for various machine classifiers \citep{2018MNRAS.474..478P}.

To fairly assess machine learning algorithms for transit detection, we recommend to perform an independent benchmark which includes {\tt BLS} and {\tt TLS}. This is beyond the scope of this paper but can be a natural follow-up work to it.

\section{Discussion}

\subsection{Arbitrary signal shapes}
{\tt TLS} can be used with arbitrary search functions to detect other kinds of periodic events in stellar (or other) light curves. We plan to implement a user-friendly interface for such functions in the next release. As an example, although stellar flares are not known to be strictly periodic, flares from TRAPPIST-1 appear to be semi-periodic \citep{2018ApJ...857...39M}. An analytic description of a stellar flare \citep{2014ApJ...797..122D} could be used to search for periodically flaring stars. What is more, many phenomena related to exoplanets were not expected or known before they were found.

It would also be possible to feed {\tt TLS} with an analytic description of exocometary transits \citep{2018MNRAS.474.1453R,2018arXiv181103102K} or disintegrating planets with comet-like tails \citep{2015ApJ...800L..21B,2018A&A...611A..63G}, atmospheric refraction \citep{2017ApJ...848...91D,2018AAS...23112802D}, exoplanetary rings \citep{2004ApJ...616.1193B,2009ApJ...690....1O,2011ApJ...743...97T,2017AJ....153..193A,2018NewA...60...88H} or artificial shapes such as rectangles \citep{2005ApJ...627..534A} as well as starshades at the Lagrange points \citep{2017MNRAS.469.4455G,2018RNAAS...2b..34M}.

\subsection{Issues with uneven sampling and data gaps}
 {\tt TLS} moves the model transit curve over the data points in phase space, very much akin to a moving window. This procedure assumes constant cadences or steps in phase space. Constant steps in phase space are only present on average, however. Transit timing variations can sometimes induce a stroboscopic effect: Cadences that are constant in time may not be constant in phase due to resonances with the observational cadence \citep{2013A&A...553A..17S}. This should only affect a small fraction of all planets.

 Variable cadences result in a morphologically distorted transit shapes, incorrect transit duration estimates, and usually reduce the SDE. Small variations of the cadence are negligible, for example from barycentering the timestamps of the Kepler satellite, which accounts for a variation of ${\lesssim}8\,$min and is therefore substantially smaller than even the shortest known transit duration of $\sim40$\,min \citep{2013ApJ...773L..15R}. Partially observed transits, which can occur near data gaps or near the beginning and end of observations, may cause similar issues. When the number of observed transits is large, for example more than a dozen, the effect of partial transits is small and this is (and will be) valid for the vast majority of planets detected by Kepler (and TESS and PLATO) due to the missions' long duty cycles. Even in the case of contamination with partial transits, the phase-folded transit light curve is usually better approximated by our {\tt TLS} default transit template than by a box.

\subsection{Observational biases due to the transit shape template}
Observational biases for transiting planets \citep{2016MNRAS.463.1323K} are partly due to the box shaped transit fit (when using BLS). Even when fitting a better transit shape like the {\tt TLS} template, similar biases can be expected since our template curve cannot be a perfect fit for all transits. For example, an eccentric or V-form grazing transit shape is substantially different from a box. This causes increased noise, resulting in lower detection efficiency. A few real, but rare, transit shapes might be closer to a box than to the reference transit template, resulting in a different set of observational biases. Characterizing these can be a natural follow-up work.

\subsection{Correlated noise}
Our definition of the SDE, extending from Eq.~\eqref{eq:chisq} to Eq.~\eqref{eq:SDE}, assumes that the noise in the light curve is uncorrelated in time. For cases of correlated noise, which is often caused by stellar activity and instrumental systematics, SDE is an overestimate. The transit evaluation metrics provided by {\tt TLS} (Appendix~\ref{sub:metrics}) include the {\tt snr\_pink} and the {\tt snr\_pink\_per\_transit}, which can be used as an indiction for correlated noise, when compared to {\tt snr} and the {\tt snr\_per\_transit}.

\begin{figure}
\includegraphics[width=\linewidth]{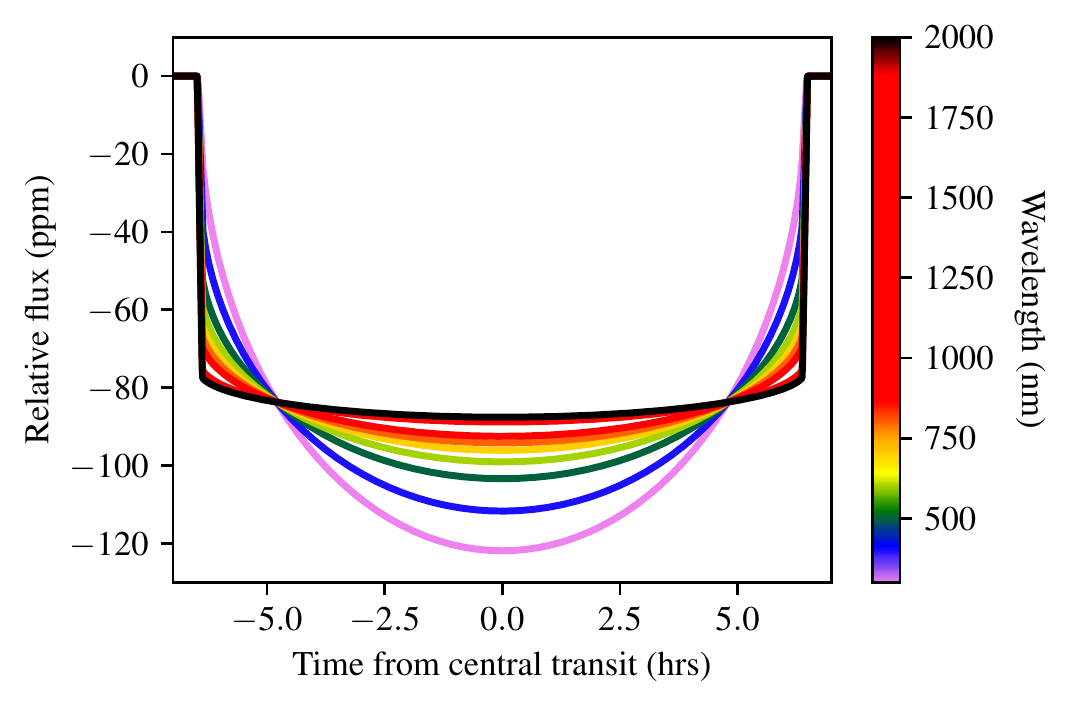}
\caption{\label{fig:wavelength}Transit light curves for different wavelengths from the UV (300\,nm) to the NIR ($2\,\mu$m) are increasingly box-shaped. Simulation for a central Earth/sun transit using measurements from \citet{1998A&A...333..338H}.}
\end{figure}

\subsection{Wavelength dependence of transit shape}

The advantage of {\tt TLS} over {\tt BLS} is largest at short wavelengths, that is in the blue-optical regime of the electronamgnetic spectrum, where stellar limb darkening is most pronounced. At longer wavelengths, transits become increasingly box-shaped and the advantage of {\tt TLS} over {\tt BLS} vanishes. As can be seen in Fig.~\ref{fig:wavelength}, a central Earth/sun transit observed at $\lambda=2\,\mu$m is well approximated by a box. While {\tt TLS} can be parameterized with appropriate limb-darkening for such a bandpass and thus transit shape, its advantage over {\tt BLS} is reduced to the trapezoidal ingress and egress shape (and the lack of binning). The detection efficiencies of {\tt BLS} and {\tt TLS} essentially match at long ($\sim2\,\mu$m) wavelengths.

\subsection{Eccentricity and transit timing/duration variations}

Throughout this paper, we have assumed planets in circular orbits with linear ephemerides. While this is a valid approximation for many real planets, a certain fraction of planets exhibits deviations in the form of eccentricity and/or transit timing/duration variations. All of these effects cause the signal to deviate from the {\tt TLS} default template and reduce the SDE. This is also true for BLS, as these effects broaden the signal in phase space. For the vast majority of these cases, the transit shapes are still closer to the {\tt TLS} default template than to a box, so that the advantage of {\tt TLS} over {\tt BLS} holds.
Usually, TTVs are very small, less than a few percent of the transit duration. Studying other cases, where a different signal template is wortwhile, could be a natural follow-up work. A search for less common transit shapes can readily be made through {\tt TLS'} interface to \textsf{batman} \citep{2015PASP..127.1161K,2015ascl.soft10002K}, using of the underlying \cite{2002ApJ...580L.171M} analytic transit model.

\subsection{Data binning}
While the time required for phase-folding and sorting is identical for {\tt TLS} and BLS, the latter is faster by an order of magnitude due to the binning of the phase-folded light curve. A box-shaped function allows for binning with minimal loss in quality and a fixed factor speed gain \citep{2014A&A...561A.138O}. For typical long cadence data (for example 30\,min), however, binning cannot be recommended for {\tt TLS} because it would smear the ingress and egress shapes of most transits and therefore reduce the sensitivity -- if the ingress and egress duration is short. For high impact parameter transits, this duration can be long, and binning may be acceptable. In general, binning is adequate if there are many data points between two phase grid points at the critical phase sampling. A detailed analysis of such trade-offs can be a natural part of follow-up work on TLS.

Our current {\tt TLS} implementation does not compensate for morphological light-curve distortions (temporal smearing effects) due to finite integration time \citep{2010MNRAS.408.1758K}. It is computationally prohibitive to re-compute the transit shape template at every test period. Instead, an optimal re-computation grid of periods could be derived following Eq.~(40) in \citet{2010MNRAS.408.1758K}. This is an open feature request for {\tt TLS} and can be part of natural follow-up work.

\section{Conclusion}
The default transit search function of {\tt TLS} is a model transit light curve optimized to find small planets. We have constructed this template based on $2\,346$ small ($R_{\rm p}/R_{\rm s}\,<\,0.05$) planets and planet candidates observed with the Kepler mission. With this template, or a user-specified signal shape, {\tt TLS} analyses the entire, unbinned data of the phase-folded light curve. Our transit injection-retrieval experiments with white noise light curves of an Earth-sized planet around a sun-like star demonstrate that these improvements yield a 17 percentage points higher true positive rate for {\tt TLS} ($\sim\,93$\,\%) compared to {\tt BLS} ($\sim\,76$\,\%) if the false alarm rates are chosen to be 1\,\%, respectively. At the same time, the {\tt TLS} false negative rate ($7\,\%$) is significantly smaller than that of {\tt BLS} ($24\,\%$). In other words, {\tt TLS} is substantially more efficient and reliable in finding small planets than {\tt BLS}.

The test statistic of {\tt TLS} is a modified version of the signal detection efficiency (SDE) used by the standard transit detection algorithm {\tt BLS}. The SDE for {\tt TLS} is derived from all the data points in the phase-folded light curves and not from the binned phase-folded light curve, as done by  {\tt BLS}. The {\tt TLS} approach is computationally more demanding but key to the increased transit detection efficiency of {\tt TLS} over {\tt BLS} for small planets. We also filter the SDE periodogram for a systematic noise component by dividing it through a walking median. The resulting SDE distribution for {\tt TLS} yields significantly more robust detections of transit-like signals compared to {\tt BLS}.

Finally, as a demonstration example for the detection of Earth-sized planets around a low-mass star, we have tested {\tt TLS} with its default transit template on the K2 data of the TRAPPIST~-~1 system and retrieved six of seven planets together with their detection statistics and phase-folded light curves. The high detection efficiency of {\tt TLS} and its optimization for computational speed makes it a natural search algorithm for small transiting planets in light curves from Kepler, K2, TESS, and PLATO.

\begin{acknowledgements}
The authors thank an anonymous referee for her or his very helpful report. RH receives funding from the German Space Agency (Deutsches Zentrum f\"ur Luft- und Raumfahrt) under PLATO Data Center grant 50OO1501. This research has made use of the Exoplanet Orbit Database, of the Exoplanet Data Explorer at \href{http://exoplanets.org}{exoplanets.org}, and of NASA's ADS Bibliographic Services. The authors made use of the following software: {\tt Astropy} \citep{2018AJ....156..123A}, {\tt NumPy} \citep{numpy:2011}, {\tt SciPy} \citep{scipy:2001}, and {\tt Matplotlib} \citep{matplotlib:2007}.
\end{acknowledgements}

\bibliographystyle{aa}
\bibliography{references}

\begin{appendix}
\label{appendix}

\section{Optimization for computational speed}
\label{sec:ap_speed}

\begin{table}
\caption{{\tt TLS} run times (in minutes) for Kepler K1 LC data, $R=R_{\odot}$, $M=M_{\odot}$, $\Delta t_0=0$}
\label{table:1}
\centering
\begin{tabular}{cccc}
\hline\hline
$\delta_{\rm cut}$ (ppm) & $\pm 0.1 \, \rho$ & $\pm 0.2 \, \rho$ & No prior \\
\hline
10  & 30 & 50 & 400 \\
50  & 20 & 40 & 300 \\
100 & 7  & 10 & 30  \\
\hline
\multicolumn{4}{l}{For reference: {\tt Astropy} {\tt BLS} run time: 13\,min}\\
\end{tabular}
\end{table}

\begin{table}
\caption{{\tt TLS} run times (in minutes) as before, but $\delta_{\rm cut}=10\,$ppm}
\label{table:2}
\centering
\begin{tabular}{cccc}
\hline\hline
$\Delta t_0$ & $\pm 0.1 \, \rho$ & $\pm 0.2 \, \rho$ & No prior \\
\hline
0     & 30 & 50 & 400 \\
0.01  & 7  & 9  & 40 \\
0.1   & 7  & 7  & 7 \\
\hline
\multicolumn{4}{l}{For reference: {\tt Astropy} {\tt BLS} run time: 13\,min}\\
\multicolumn{4}{l}{$\Delta t_0$ is the step in phase space in units of transit duration}\\
\end{tabular}
\end{table}

A straightforward implementation of the mathematical framework described in Sect.~\ref{sec:methods} into an algorithm would result in huge computational demands with days of run time per K2 light curve for a reasonably dense grid of $P, t_0, \delta, d$ trial values. We thus implement a computationally optimized version that produces identical results to a straightforward coding. Most aspects of our optimization are time-memory trade-offs, where repetitive calculations are identified and stored in memory after their first calculation. A memory read is often faster than a repeated calculation. The memory size requirements for the computer's random access memory (RAM) are of order 50\,MB per thread, dominated by the buffered signal shapes.

Most of the data points in a light curve are out of transit. As a consequence, the out-of-transit data account for the majority of computation time. Most important, the out-of-transit data of both the observed and of the modeled light curves are identical for a fixed trial epoch, transit duration, and orbital period. Hence, we design {\tt TLS} to calculate the squared residual values of the out-of-transit data only once in the ($P$-$t_0$-$d$) 3D parameter hyperspace of our 4D search grid. Consequently, only a small amount of in-transit squared residuals need to be calculated for the various trial depths of the transits.

We also found that instead of calculating the oversampled transit model for each $\chi^2$ test in the $d$ space, it is significantly faster to pre-compute all oversampled transit models for all trial durations but only for single, arbitrary transit depth for any given $d$ value. Re-scaling the transit in $\delta$ only requires one multiplication per data point. Resampling a model in width, however, would be considerably more expensive due to the necessary oversampling. Thus, the transit models are pre-computed and cached for any given transit duration and then they are scaled in depth on-the-fly as {\tt TLS} searches through the transit duration grid.

Phase-folding involves sorting the orbital phases $\phi_i\,=\,t_i/P$, where $t_i$ are the times of observations. Sorting can be extremely demanding computationally. A typical K2 light curve worth 4000 data points takes 0.5\,ms to fold and sort on an Intel Core i7-7700K. The required time for folding and sorting 20\,000 trial periods and 4000 trial phases would be $\sim\,11\,$hrs. Hence, as in several {\tt BLS} implementations, {\tt TLS} only folds and sorts the phases once per trial period, which takes $\sim\,5\,$sec and implies a speed gain of a factor of 4000. We find that the fastest algorithm to sort phase arrays is ``Mergesort'' (von Neumann 1945, unpublished), which is typically $\sim\,20\,$\% faster than the commonly used Quicksort algorithm \citep{Hoare1962}.

The {\tt TLS} reference implementation is written in pure {\tt python} code, which is an interpreter-based programming language and thus comes with a speed loss. We therefore chose to implement many of the time-critical parts of {\tt TLS} with the specialized {\tt numba} package \citep{Lam2015} that translates the {\tt python} code in machine code. This procedure is called ``just-in-time'' compilation and saves us two orders of magnitude in computing time.

{\tt TLS} can be adjusted for adequate performance on large data sets with minor compromises. When speed is critical, we recommend to first perform a search using fast binned BLS, which is adequate to recover all high-S/N planets. Using iterative runs, these significant signals can be found and masked from the data, reducing overall variance and data volume. When no significant {\tt BLS} signal remains, the search can be switched to TLS.

The computational speed of {\tt TLS} can then be increased with a sensible threshold for the shallowest transit signal that is fitted. For example, instrumental and stellar noise limit the detectability of signals by Kepler to transit depths of $\sim100\,$ppm for periods of $\sim365\,$d around a G2V-star, roughly the Earth-equivalent that the Kepler mission was originally designed for. Allowing the algorithm to fit signals of down to 10\,ppm will then only result in a very high computational load, but almost certainly not in the discovery of a real 10\,ppm transit signal. As the S/N of a planet is a complex combination of many factors, we have set the default {\tt TLS} parametrization to a threshold of 10\,ppm. For reference, the shallowest known transit is 11.9 ppm \citep[Kepler-37\,b,][]{2013Natur.494..452B}, a discovery that was made possible due to the short orbital period (13.4\,d) and the stacking of $\sim100$ transits. For long-period planets, one may choose thresholds of, for example, 50\,ppm for Kepler data or 100\,ppm for K2 data, which avoids fitting out the complete noise floor. For reference, an Earth-sized planet transiting a dG2 star has a transit depth of $\sim100\,$ppm. Then {\tt TLS} run times are similar to those of {\tt BLS}. As a follow-up work, one might develop a heuristic of a shallowest transit depth to be fitted, as a function of noise in the data, period, and other factors.

For Kepler K1 LC data (30\,min cadence over 4.25\,yr), we have tested {\tt TLS} run times for different combinations of stellar density priors (from stellar radii and masses), shallowest transit fit depths ($\delta_{\rm cut}$), and phase space sampling $\Delta t_0$ (Tables~\ref{table:1},~\ref{table:2}). The KIC and EPIC catalogs \citep{2011AJ....142..112B, 2016ApJS..224....2H} typically have relative mass and radius uncertainties of 5\,\%, so that a range of $\pm 0.1 \, \rho$ ($\pm 0.2 \, \rho$) gives a $2\,\sigma$ ($4\,\sigma$) confidence interval. Priors decrease run times by an order of magnitude in case of complete phase space sampling. Then, the influence of $\delta_{\rm cut}$ is a factor of a few between 100\,ppm and 10\,ppm for typical K1 data, where the standard deviation per data point (after detrending) is typically in the hundreds of ppm.

{\tt TLS} also offers the option of not sampling the phase space at every cadence. For example, K1 LC data (60000 points) allows for transit signal lengths of up to 7200 points for $T_{14}/P=0.12$. Considering noise levels in the hundreds of ppm per point, shifting this long template point by point is pointless. Instead, {\tt TLS} can shift the data in phase space by a user-defined fraction of the transit duration, the latter of which is measured in cadences. As an example, the default value of 1\,\% shifts a transit signal of 200 points length by 2 points in each trial, saving 50\,\% of the computational time. This procedure results in much faster run times because most of the computational effort is spent to test very long transit duration. Empirically, we find that setting $\Delta t_0=0.01$ (instead of zero) allows for virtually identical SDE values in Kepler K1 and K2 data, while $\Delta t_0=0.1$ results in a detection efficiency loss of a few percent and should be used only for high S/N transits, for example, using an iterative search.

Our {\tt TLS} implementation leverages all available CPU cores and shows continuous updates of the estimated remaining time and a progress bar. The user can use the estimate to balance run time and search depth.

\section{Edge effect jitter in {\tt BLS} which leads to additional noise}
\label{sub:edge}

\begin{figure}
\includegraphics[width=\linewidth]{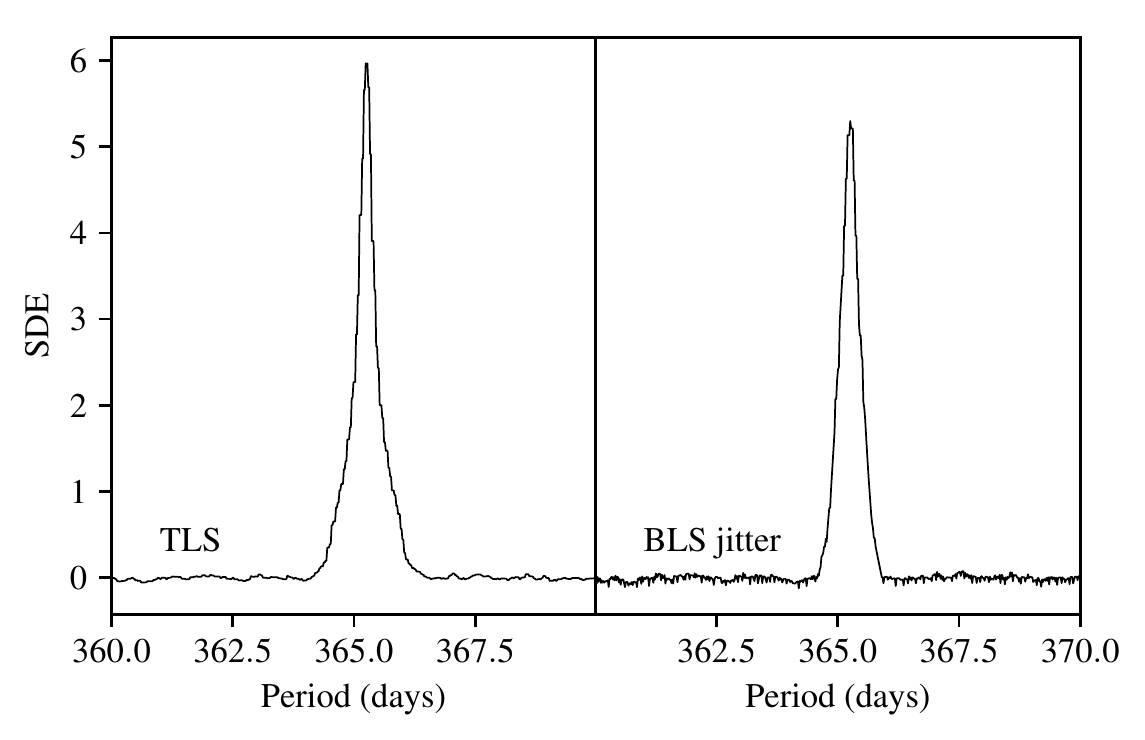}
\caption{\label{fig:edge_effect_jitter}Edge effect jitter in {\tt BLS} (right panel) and absence of this jitter in {\tt TLS} (left).}
\end{figure}

The original {\tt BLS} implementation did not account for transit events occurring to be divided between the first and the last bin of the folded light curve. This was noted by Peter R. McCullough in 2002, and an updated version of {\tt BLS} was made (\textsf{ee-bls.f}) to account for this edge effect. The patch is commonly realized by extending the phase array through appending the first bin once again at the end, so that a split transit is stitched together, and present once in full length. The disadvantage of this approach has apparently been ignored: The test statistic is affected by a small amount of additional noise. Depending on the trial period, a transit signal (if present) is sometimes partly located in the first and the second bin. The lower (in-transit) flux values from the first bin are appended at the end of the data, resulting in a change of the ratio between out-of-transit and in-transit flux. There are phase-folded periods with one, two, or more than two bins which contain the in-transit flux. This causes a variation (over periods) of the summed noise floor, resulting in additional jitter in the test statistic. For typical Kepler light curves, the reduction in detection efficiency is comparable to a reduction in transit depth of $\sim 0.1-1\,$\%. {\tt TLS} corrects this effect by subtracting the difference of the summed residuals between the patched and the non-patched phased data. A visualization of this effect on the statistic is shown in Fig.~\ref{fig:edge_effect_jitter}, using synthetic data. In real data, the effect is usually overpowered by noise, and was thus ignored, but it is nonetheless present.

\section{Transit evaluation metrics}
\label{sub:metrics}

In addition to the SR and SDE transit search statistic (Sect.~\ref{sec:transit_search}), our {\tt python} implementation of {\tt TLS} outputs the period of the highest SDE value, its corresponding $t_0$, $\delta$, $d$, and the planet-to-star radius ratio for zero transit impact parameter $R_P/R_\star~=~\sqrt{\delta (1 - c_1/3 - c_2/6)}$ \citep{Heller2018_LD}, where $c_1$ and $c_2$ are the limb darkening coefficients of a quadratic limb darkening law. {\tt TLS} also offers a range of automated evaluation parameters of the detected transits such as
\begin{itemize}
\setlength\itemsep{-0.1em}
\item the ratio of the signal to the white noise of the stacked transits ({\tt snr})
\item the ratio of the signal to the pink noise of the stacked transits ({\tt snr\_pink}) \citep{2006MNRAS.373..231P,2016A&C....17....1H}
\item the ratio of the signal to the white noise of the individual transits ({\tt snr\_per\_transit})
\item the ratio of the signal to the white noise of the individual transits ({\tt snr\_pink\_per\_transit})
\item the significance (in units of standard deviations) between the depths of the odd and even transits ({\tt odd\_even\_mismatch})
\item the number of transits with in-transit data points ({\tt distinct\_transit\_count})
\item the number of transits with no in-transit data points ({\tt empty\_transit\_count})
\item the total number of transits ({\tt transit count})
\item the number of data points for each transit ({\tt per\_transit\_count})
\end{itemize}
\noindent

Our online release of {\tt TLS} comes with a documentation of these parameters and of additional evaluation metrics.

\end{appendix}
\end{document}